\documentclass[10pt]{iopart}
\usepackage{graphicx}
\usepackage{amsmath}
\usepackage[normalem]{ulem}
\usepackage{color,soul}
\newcommand{\stkout}[1]{\ifmmode\text{\sout{\ensuremath{#1}}}\else\sout{#1}\fi}

\def\hlMode{0} % 1 = highlight on; 0 = highlight off
\newcommand{\hlcolor}[2]{\sethlcolor{#1}\hl{#2}}

\newcommand{\hb}[1]{\hlcolor{cyan}{#1}}

\newcommand{\cbox}[2]{\fcolorbox{#1}{white}{#2}}

\newcommand{\blbox}[1]{\cbox{cyan}{#1}}

\if\hlMode0
    \renewcommand{\hlcolor}[2]{#2}
    \renewcommand{\cbox}[2]{#2}
\fi

\begin{document}

\title{Exploration of Non-Resonant Divertor Features on the Compact Toroidal Hybrid}

\author{K. A. Garcia$^1$, A. Bader$^{1,3}$, H. Frerichs$^1$, G. J. Hartwell$^2$, J. C. Schmitt$^{2, 3}$, N. Allen$^2$ and O. Schmitz$^1$}

\address{$^1$ Department of Nuclear Engineering \& Engineering Physics, University of Wisconsin - Madison}
\address{$^2$ Department of Physics, Auburn University}
\address{$^3$ Type One Energy, Madison Wisconsin}
\ead{kgarcia26@wisc.edu}

\begin{abstract}

Non-resonant divertors (NRDs) separate the confined plasma from the surrounding plasma facing components (PFCs). The resulting striking field line intersection pattern on these PFCs is insensitive to plasma equilibrium effects. However, a complex scrape-off layer (SOL), created by chaotic magnetic topology in the plasma edge, connects the core plasma to the PFCs through varying magnetic flux tubes. The Compact Toroidal Hybrid (CTH) serves as a test-bed to study this by scanning across its inductive current. Simulations observe a significant change of the chaotic edge structure and an effective distance between the confined plasma and the instrumented wall targets. The intersection pattern is observed to be a narrow helical band, which we claim is a resilient strike line pattern. However, signatures of finger-like structures, defined as heteroclinic tangles in chaotic domains, within the plasma edge connect the island chains to this resilient pattern. The dominant connection length field lines intersecting the targets are observed via heat flux modelling with EMC3-EIRENE. At low inductive current levels, the excursion of the field lines resembles a limited plasma wall scenario. At high currents, a private flux region is created in the area where the helical strike line pattern splits into two bands. These bands are divertor legs with distinct SOL parallel particle flow channels. The results demonstrate the NRD strike line pattern resiliency within CTH, but also show the underlying chaotic edge structure determining if the configuration is diverted or limited. This work supports future design efforts for a mechanical structure for the NRD.

\end{abstract}
\noindent{Keywords\/}: stellarator, divertor, modeling \\
%\submitto{\NF}
\maketitle

\section{Introduction}
\label{section:intro}
The non-resonant divertor (NRD) is an alternative divertor solution for stellarators to the island divertor concept explored at W7-X or the closed helical divertor used at LHD \cite{boozer_stellarator_2015, boozer_simulation_2018, punjabi_simulation_2020}. The name arises from the fact that the NRD is not reliant on having a specific rotational transform, $\stkout{\iota} = \iota / 2\pi$, value in the edge \footnote{The rotational transform is defined as $\stkout{\iota} = d \psi / d \Phi $ where $\psi $ is the poloidal magnetic flux and $\Phi$ is the toroidal magnetic flux. It is the ratio of poloidal transits to toroidal transits for a field line.}. This is in contrast to resonant divertors which exploit the resonance in the rotational transform to produce an island, which is used for the divertor structure in e.g. the island divertor explored at W7-AS and W7-X \cite{grigull_first_2001, renner_divertor_2002, wolf_performance_2019}. 

The outlook for NRDs is important for the future of stellarator-based fusion energy. The need for stable stellarator divertor concepts under equilibrium changes arises because there are promising optimized stellarator configurations that require significant equilibrium changes as plasma pressure is built up. This is especially true for quasihelically symmetric (QHS) and quasiaxisymmetric (QAS) devices, which will produce self-generated bootstrap currents at finite plasma pressure \cite{nuhrenberg_quasi-helically_1988, landreman_optimization_2022}. These currents depend on the pressure profile, which in turn depends on the plasma transport, and therefore, it is hard to predict except in idealized cases \cite{redl_new_2021}. Resonant divertors are very sensitive to the rotational transform profile, and small errors in the predicted current could move the islands so that heat flux impinges on undesired regions of the wall. If the requirement to fix the rotational transform profile is removed, then there is significantly more configurational flexibility available. 

Previous work has shown that in NRDs, the overall magnetic structure on wall surfaces \cite{punjabi_simulation_2020} and also the deposited heat flux on the wall is resilient to changes in the plasma equilibrium, specifically to changes in the shape of the last closed flux surface (LCFS) \cite{bader_hsx_2017}. The resilient feature of the NRD concept can be categorized into (a) features related to the magnetic structure and shape of the plasma boundary and (b) features of the intersection pattern of escaping field lines on the wall, also known as the strike point or strike line pattern, which are indicative of the plasma flux on the PFCs and on the divertor target. If these features are maintained under a given plasma equilibrium change, we call the scenario resilient.\hb{ It should be noted that an objective definition for resiliency does not exist. The amount of allowed variation of the plasma-wall interaction region depends on details of the divertor structure in question. }   

\hb{Resiliency of these categorical features were seen in previous results if 1) the LCFS was not limited directly by the wall and 2) large islands did not exist in the plasma edge.} By moving the PFCs radially outwards in a given configuration, it is possible to transition from a limited configuration to a non-limited one. In the non-limited case, a radial gap between the confined plasma and the wall exists which is filled with a combination of open field lines, small islands, cantori and stochastic regions \cite{punjabi_magnetic_2022}. Many of these features have been explored in mathematically idealized equilibria \cite{helander_stellarator_2012, boozer_simulation_2018, punjabi_simulation_2020}. In this paper, we will discuss the structures that exist in the edge in a series of configurations where these structures vary significantly, and we will also examine their evolution. Topics such as resiliency and the transition between limited and diverted scenarios will be discussed in the process of describing this evolution. 

We focus on understanding the coupling of the magnetic structure in the plasma boundary to the overall resilient intersection pattern on wall elements and the consequent heat deposition features. This is done in tandem with understanding the basic divertor and boundary transport aspects that form these divertor flux patterns. Previous research into these aspects have been performed on a limited number of devices including HSX \cite{bader_hsx_2017}, W7-X \cite{strumberger_magnetic_1992} and for a very limited configuration set in CTH \cite{bader_minimum_2018}.

We expand on previous results of CTH to explore the edge structures in more detail. Specifically, through configuration variation of the induced plasma current driven by the solenoid of CTH, it is possible to change the rotational transform over a broad range and hence transform the edge behavior strongly. It is shown for the first time that a chaotic edge structure is formed from these configuration changes and yet, these intersection patterns on the wall elements are resilient to the edge structure evolution. \hb{The structural details of the chaotic magnetic boundary will be identified from the intersection patterns on several wall targets in section \mbox{\ref{section:flare}}.}. It is shown that these details in the intersection pattern dictate the heat flux pattern and that the resulting heat flux is shifted within the overall resilient helical intersection pattern across these configuration changes. The implications of these findings for possible mechanical divertor designs are discussed. 

In section \ref{section:cthedge}, the CTH experiment is described and the set of configurations that will be used for the rest of the paper is introduced. In section \ref{section:flare}, the calculations of the strike line evolution as the configuration changes are presented and they are compared to connection length analysis on the wall surfaces at different radii. In section \ref{section:emc3}, EMC3-EIRENE simulations for dedicated configurations are provided. In section \ref{section:disc}, the results are discussed in terms of the impact of designing mechanical structures around the magnetic structure of the boundary in an NRD configuration. Conclusions are also provided in section \ref{section:disc}.

\section{CTH Edge Structure}
\label{section:cthedge}

\begin{figure}
    \centering
    \includegraphics[scale=0.15]{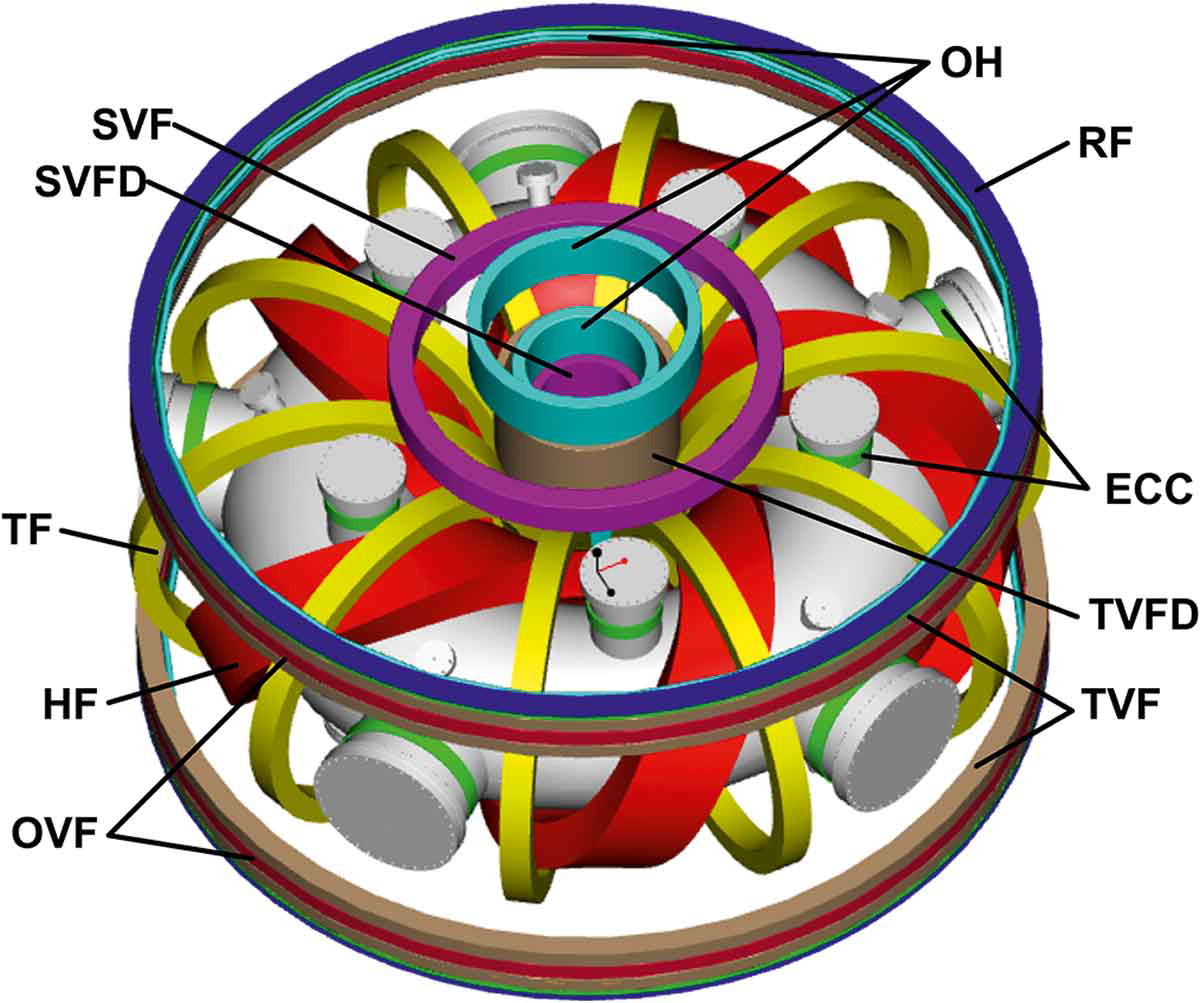}
    \caption{Vacuum vessel and magnetic coils of CTH \cite{hartwell_design_2017}. }
    \label{fig:cth_coils}
\end{figure}

The Compact Toroidal Hybrid (CTH), depicted in figure \ref{fig:cth_coils}, is a 5-field period torsatron device that also features a central solenoid that can be used to inductively drive toroidal plasma current $I_p$ \cite{peterson_initial_2007}. The ``hybrid" part of the name refers to the fact that CTH spans the space between a tokamak and a stellarator. The vacuum vessel is a circular torus with major radius $R_0=75\ \text{cm}$ and minor radius $a=29\ \text{cm}$.  A CAD rendering of the CTH layout is shown in figure \ref{fig:cth_coils} from \cite{hartwell_design_2017}. In this figure, the helical coil that produces the 3D field is shown in red and the central solenoid is shown in brown. CTH includes many shaping coils, but in this paper, the current in all the coils is kept fixed, and the only difference between configurations is the amount of $I_p$ driven in the plasma by the central solenoid. The set of coil currents for the configuration considered here is shown in \hbox{Table \ref{tab:cth_coilset}}. 

\subsection{\hb{Generating Equilibria for Analysis}}

The systematic exploration of the NRD at CTH is conducted by an analysis of the magnetic field structure in the plasma edge on various wall positions for six different levels of $I_p$, ranging from no inductive current to \hbox{$10\ \text{kA}$}. \hb{The centrally-peaked current density profile is proportional to \hbox{$I'_p = d I_p/ds \propto \left(1-s^3\right)^5$}, where $s$ is the normalized toroidal magnetic flux. This current profile is similar to the 2-power parameterization that is empirically motivated by CTH experimental data \mbox{\cite{ma_determination_2018}}.} A Variational Moments Equilibrium Code (VMEC) simulation \cite{hirshman_momcon_1986} in free-boundary mode establishes an MHD equilibrium solution for each of the six levels of $I_p$. The ohmic current drive varies the rotational transform as seen in figure \ref{fig:iotabar}. 

\hb{CTH can be scanned over a range of rotational transforms, and in this paper, the chosen $I_p$ were expected to produce non-limited plasma wall configurations at the higher $I_p$ values. In figure \mbox{\ref{fig:iotabar}}, the rotational transform} $\stkout{\iota}$ as a function of major radius $R$ is shown for six equidistant values of $I_p$. The top plot is generated from field line following and calculating $\stkout{\iota}$ while the bottom plot is the profile generated from the VMEC solution. \hb{It is noted that each $I_p$ has a different magnetic axis and for each $I_p$ case the axis shifts roughly between $ 0.71 \text{ cm } < R < 0.79 \text{ cm }$, and this can be seen in figure \mbox{\ref{fig:poincare}}}. From these plots \hb{in figure \mbox{\ref{fig:iotabar}}}, it can be seen that the configurations feature a very low magnetic shear, with almost entirely flat rotational transform profiles and some variation towards the plasma edge. We also observe a shift of the profile from around $0.3-0.35$ for no current drive, to $0.7-0.8$ at $10$ kA as $I_p$ is increased. The flat profiles and the different values of $\stkout{\iota}$ imply that the equilibrium as a whole, and in particular the plasma edge, is governed by very different rational surfaces for each value of $I_p$. 

\hb{In each of the plots of figure \mbox{\ref{fig:iotabar}}}, these possible rational surfaces are shown in dashed black lines for the rotational transforms values of $\stkout{\iota} = \frac{5}{6}, \ \frac{5}{7}, \ \frac{5}{8}, \ \frac{5}{9}, \ \frac{5}{10}, \text{ and } \frac{5}{11}$. As $I_p$ is increased, we see that each current case coincides with a particular $\stkout{\iota}$ and eventually interacts with another separate set of rational surfaces. For example, in the case of $I_p= 4 $ kA, shown as a blue profile, this profile is governed by the $\stkout{\iota}=\frac{5}{10}$ island chain but eventually lowers to $\stkout{\iota}=\frac{5}{11}$ towards higher $R$ at the edge. The impact of the magnetic structure is accordingly expected to be significant and, therefore, enables us to study how much the field line intersection patterns on wall elements are resilient to such changes in the equilibrium. 

The next step to prepare the field line tracing is to perform a virtual casing calculation on the VMEC equilibrium to determine the plasma contribution to the field outside the LCFS. The magnetic potential from the virtual casing calculation is added to the result from a Biot-Savart calculation from the coils themselves, producing a continuous vector potential, $\Vec{A}_{Total}=\Vec{A}_{Plasma}+\Vec{A}_{Coils}$. The magnetic field is derived from the relation \hbox{$\nabla \times \vec{A}_{Total} = \vec{B}$}. This process ensures a divergence-free magnetic field throughout all space. These steps are carried out by the BMW code \cite{cianciosa_notitle_2016}. 

 \subsection{\hb{Calculating the Magnetic Field Structure}}

\begin{figure*}
    \centering
    \includegraphics[scale=0.8]{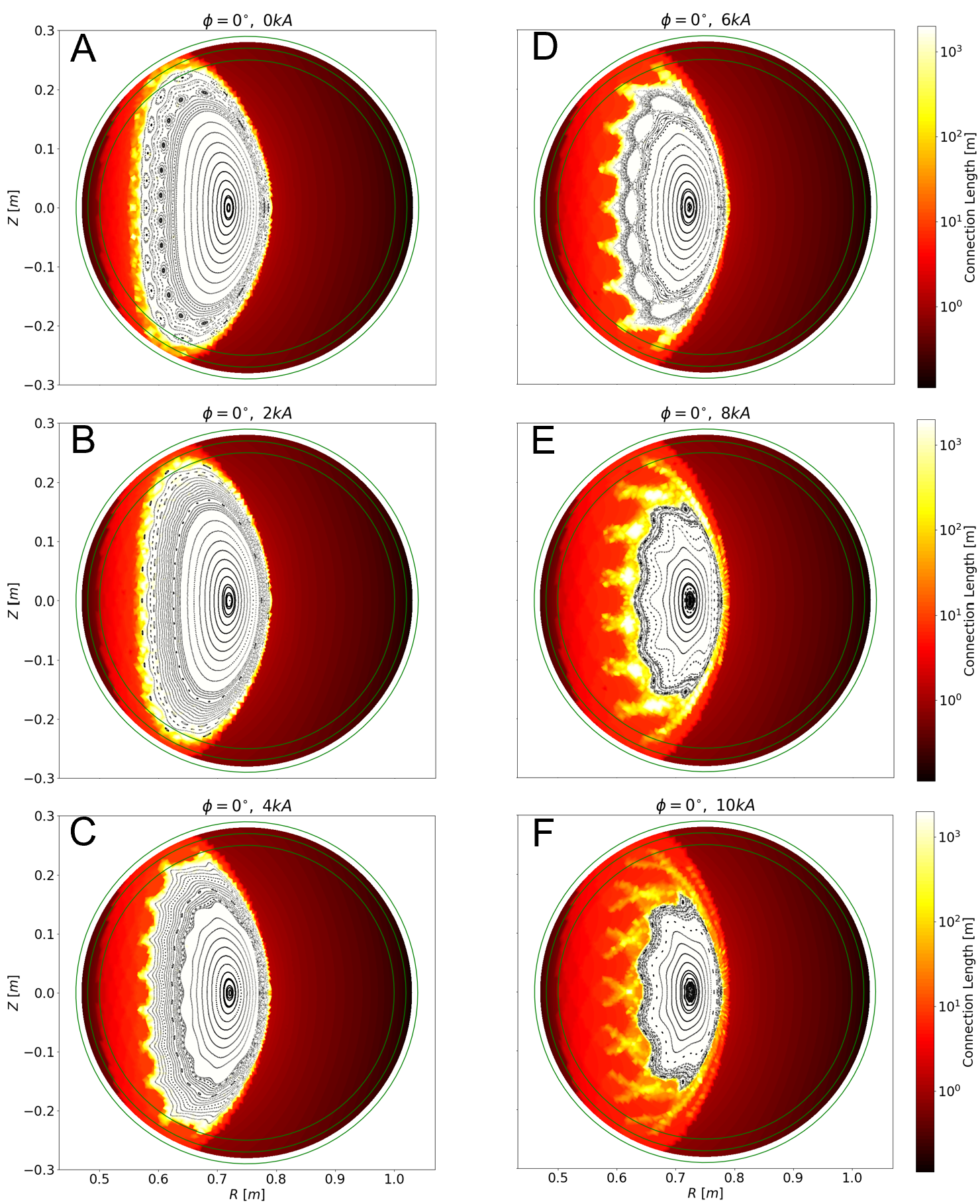}
    \caption{Logarithmic connection length ($Lc$) contour map for $I_p \in [0 \text{ kA}, 10 \text{ kA}]$ for $\phi=0^{\circ}$. A Poincar\'e map is superimposed as black points on all plots. Additionally, three circles are shown in green for $r=25$ cm, $r=27$ cm, and $r=29$ cm to indicate the locations of the radial wall targets in the analysis performed in sections \ref{section:flare} and \ref{section:emc3}. }
    \label{fig:poincare}
\end{figure*}

With this, the magnetic field structure can be calculated using the Field Line Analysis and Reconstruction Environment (FLARE) code \cite{frerichs_notitle_2015}. FLARE requires a geometric boundary representation, and for this, we use the circular vessel wall of CTH and we do not consider any substructures in the vessel wall. The field line structure for the configurations can be seen, for the $\phi = 0$ plane, in figure \ref{fig:poincare}. In this figure, an overlay of the Poincar\'e map in black and the magnetic field line connection length $L_c$ as color-coded contour map is shown. The six different plots are for an increasing level of inductive current, starting at $0$ kA in the upper left corner in figure \ref{fig:poincare} A and increasing in $2$ kA increments to $10$ kA in the plot labeled F. The field line tracing calculation is performed for a maximum connection length $L_c$ of $1$ km in both the forward and backward toroidal direction along the field line. The white surfaces represent the confined plasma, that is, locations where $L_c > 1\ \text{km}$ in both directions. This is why the maximum combined $L_c$ is $2$ km on the color bar as it is the sum of both toroidal directions. The field line tracing calculation shown here is performed with respect to the boundary at $29\ \text{cm}$. 

\begin{table}
    \centering
    \begin{tabular}{c|c|c}
         Coil ID& \# of Windings (Polarity)&Current (A) \\
         \hline
        HF$^*$ & 4 & 3333 \\
        OVF$^*$ & -6 & 3333 \\
        TVF & -4 & 535 \\
        SVF &30 & -200 \\
        TF &48 & 900 \\
        \hline
    \end{tabular}
    \caption{CTH coilset ID, number of turns in the coil, and currents for high- $\stkout{\iota}$ operation.  If an ID is not listed, it carries no current. $^*$Connected in series.}
    \label{tab:cth_coilset}
\end{table}

\begin{figure}
    \centering
    \blbox{\includegraphics[scale=0.35]{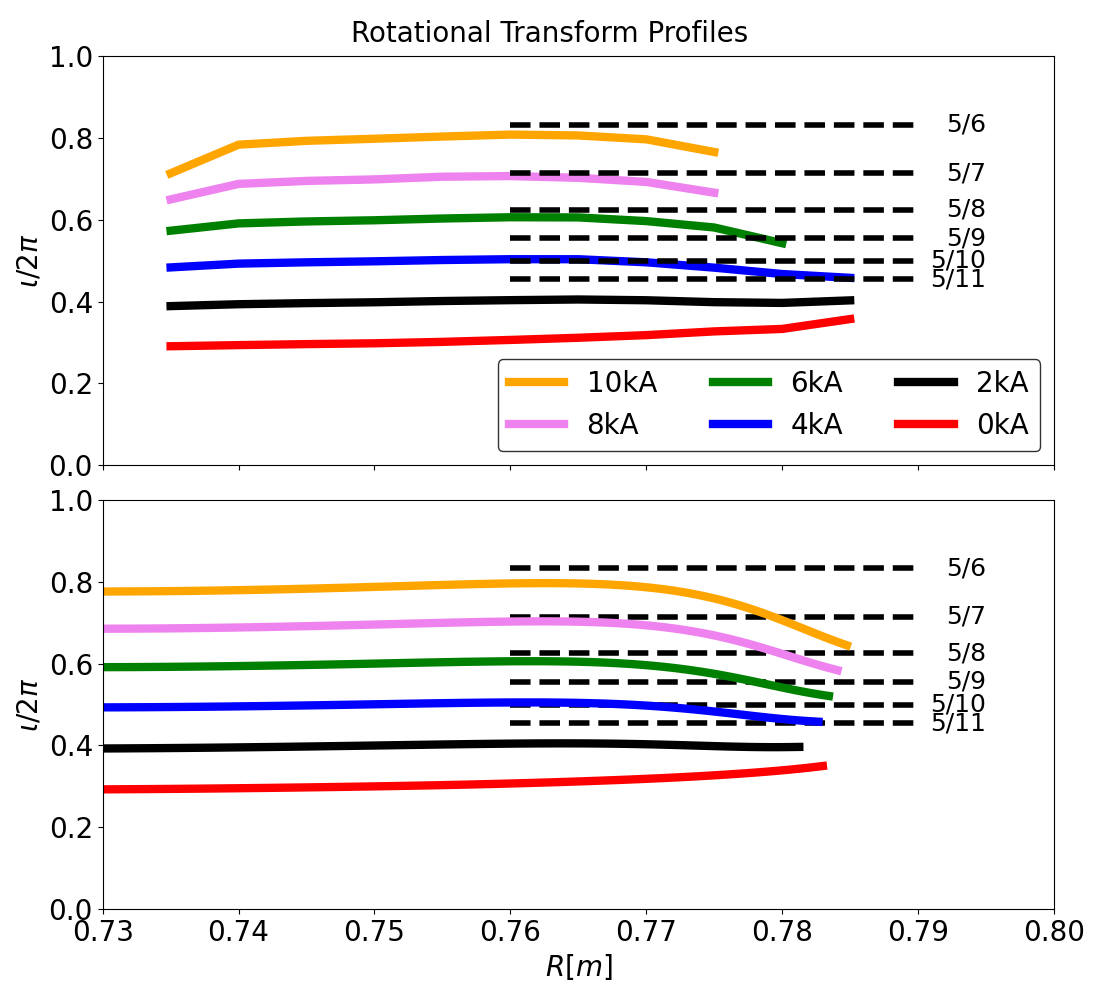}}
    \caption{Rotational transform profiles $\stkout{\iota}$ for $I_p \in [0kA,10kA]$ as a function of $R$. Top plot is calculated via field line following while the bottom plot is generated from the VMEC solution. \hb{Both plots share the same legend.}}
    \label{fig:iotabar}
\end{figure}

The Poincar\'e plots allow us to identify characteristic domains of the magnetic structure in the boundary. The inner plasma core is seen as long field lines (white color) forming intact magnetic flux surfaces. More radially outward of the core, the presence of magnetic island chains can be seen in all configurations. In the $0$ kA case, the islands are well formed. However, as $I_p$ is increased, the $\stkout{\iota}$ profile rises and lower order poloidal harmonics define the island structure. Because they feature a larger radial width, adjacent island chains start to overlap and a chaotic region is formed. The structure of this chaotic edge is characterized by a mix of smaller and larger $L_c$ flux channels formed by this interaction of magnetic islands from adjacent rational surfaces. The chaotic domain has areas of finite $L_c$ which shows that this domain is connected to the wall elements on the maximum length scale of the field line tracing. Therefore, we call this yellow to red colored domain in figure \ref{fig:poincare} an open chaotic layer. The dark red to black areas with very short connection lengths, $L_c < 1\ \text{m}$ is the region far outside of the LCFS and represents the unperturbed vacuum magnetic field far away from the plasma boundary. The confined core region, seen as white colored field domain with intact magnetic flux surfaces, is observed to shrink significantly with increasing $I_p$ and the rendering of the closed field lines become increasingly corrugated at the plasma edge. This is most visible in the right plots in figure \ref{fig:poincare} D$-$F with currents from $6-10$ kA. This corrugation inhibits calculation of the $\stkout{\iota}$ values for higher $R$ as discussed for the left plot of figure \ref{fig:iotabar}. The corrugated features of the magnetic field lines are characterized by magnetic flux bundle structures of finite $L_c$ that resemble the shape of fingers. These fingers structures grow radially outward and tend to intersect one another at the highest values of $I_p$. A key question that presents itself looking at these strong structural changes with increasing $I_p$ is how the significant changes in chaotic magnetic boundary structure couple to the intersection of this open chaotic layer with the wall surfaces. 

\begin{figure}
    \centering
    \includegraphics[scale=0.225]{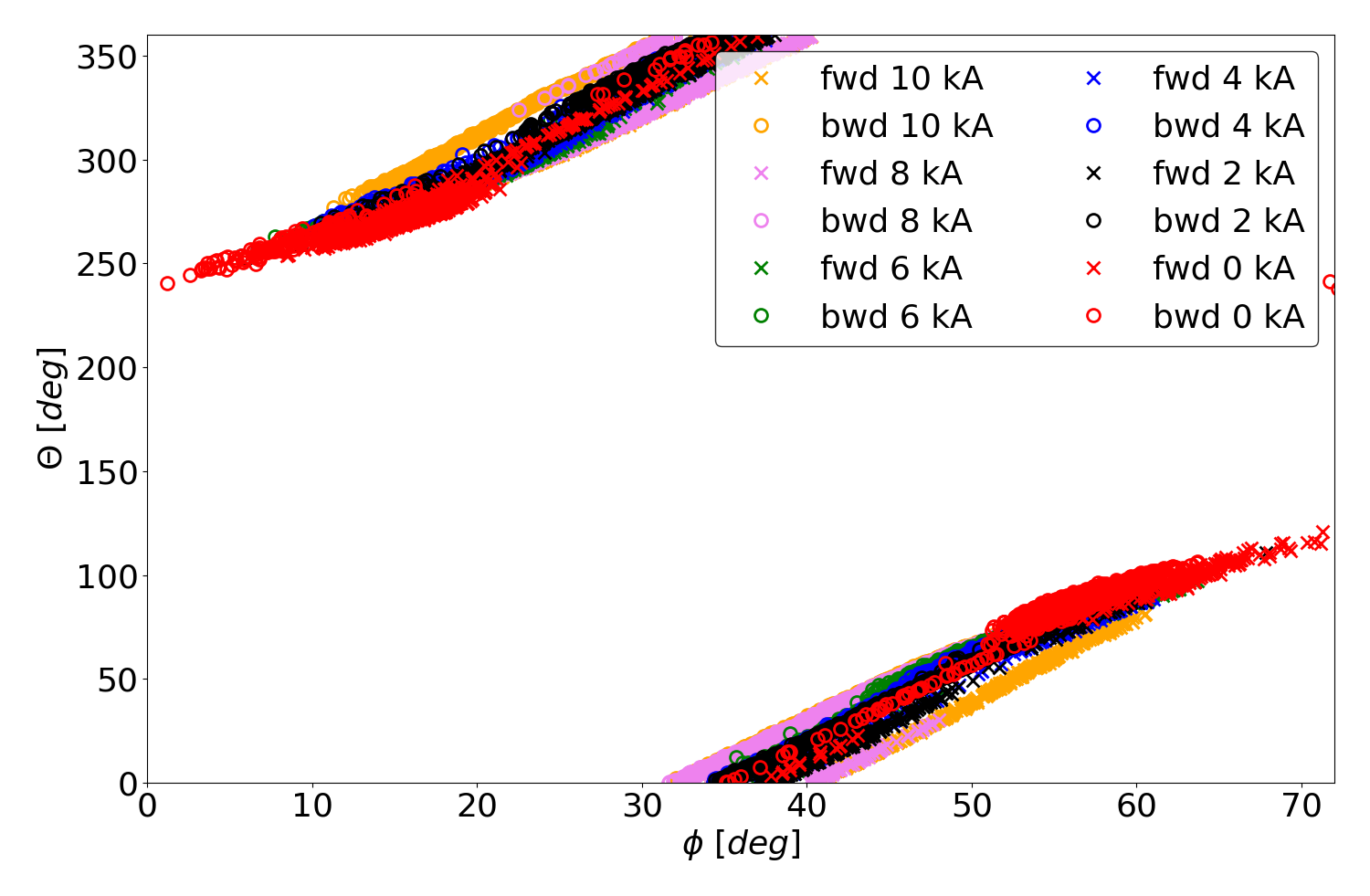}
    \includegraphics[scale=0.225]{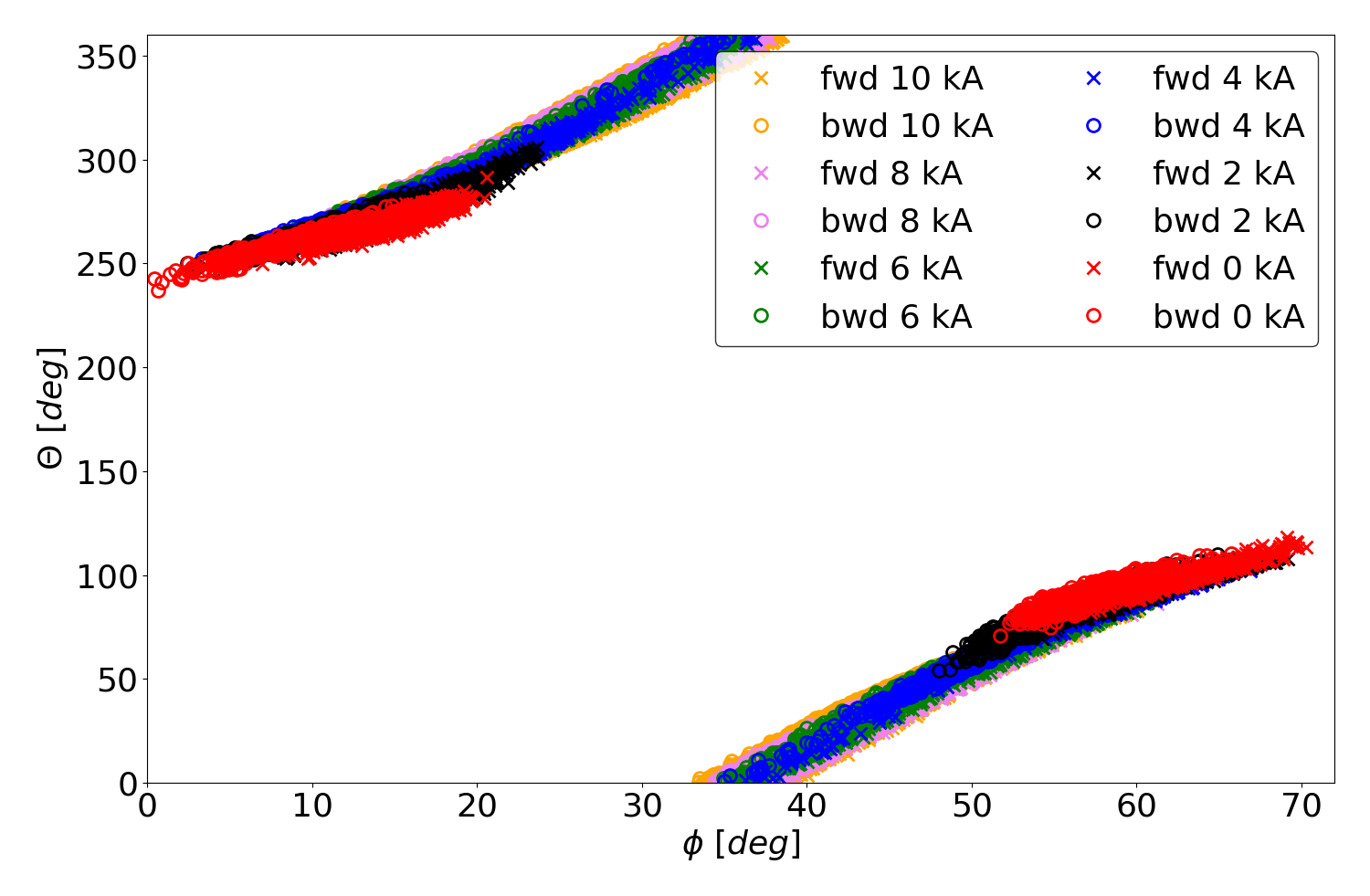}
    \caption{Forward and backward strike point locations along the CTH boundary at $r=29$ cm (top) and $r=25$ cm (bottom) for $d=4.4 \times 10^{-7} \text{m}^2/\text{m} $ followed for a maximum $L_c$ of $10$ km plotted for a single field period. }
    \label{fig:spoints}
\end{figure}

\begin{figure*}
    \centering
    \includegraphics[scale=0.68]{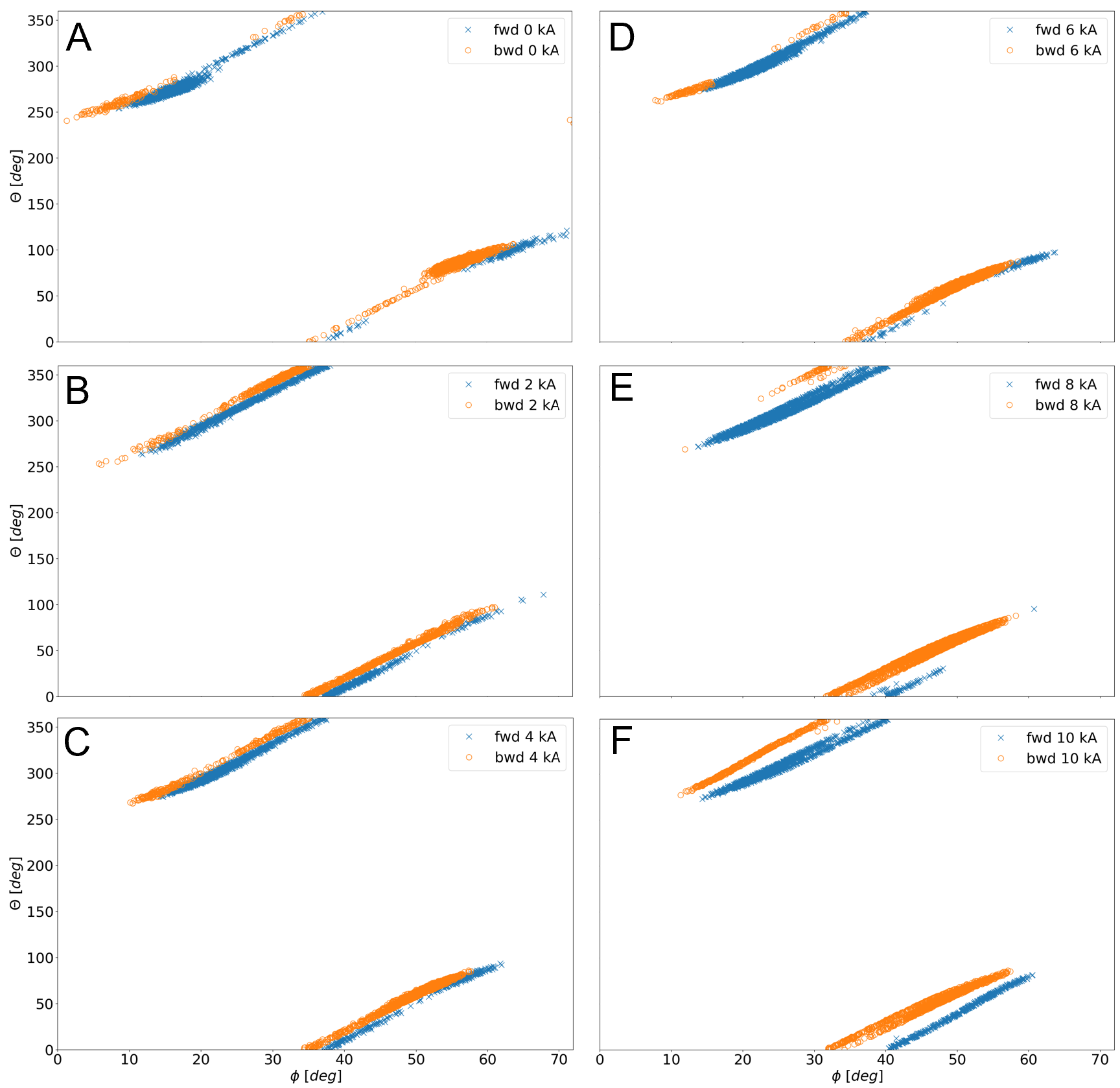}
    \caption{Forward and backward strike points for $r=29cm$ plotted for all $I_p$ cases for $D=4.4 \times 10^{-7} \text{m}^2/\text{m} $ followed for a maximum $L_c$ of $10$ km plotted for a single field period.}
    \label{fig:spoints29}
\end{figure*}

\begin{figure*}
    \centering
    \includegraphics[scale=0.68]{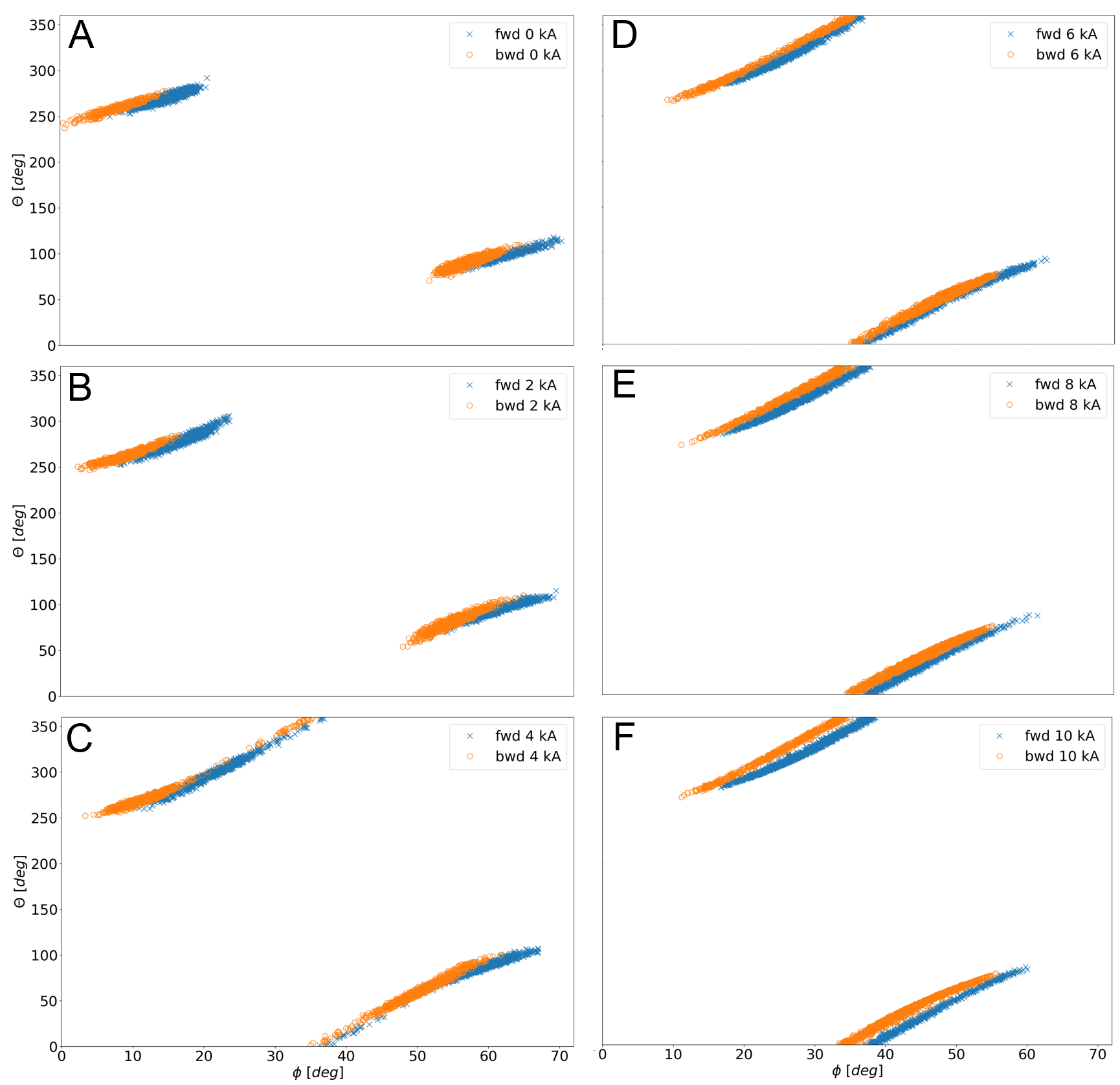}
    \caption{Forward and backward strike points for $r=25$ cm plotted for all $I_p$ cases for $D=4.4 \times 10^{-7} \text{m}^2/\text{m} $ followed for a maximum $L_c$ of $10$ km plotted for a single field period.}
    \label{fig:spoints25}
\end{figure*}

\subsection{\hb{Description of the Chaotic Layer}}

\hb{In the vicinity of a magnetic island, there will be an island separatrix that is defined by the set of field lines passing by asymptotically to the island X-points. The X-points are defined by two intersecting manifolds, which represent the trajectories of these asymptotically approaching field lines. These are generally referred to as unstable and stable manifolds. These manifolds can have lobe-shaped, or finger-like, trajectories, when projected onto one poloidal plane. The path created by overlapping stable and unstable manifolds of one single island chain is referred to as a homoclinic tangle. In the case where the stable and unstable manifolds from different island chains intersect, the resulting path is referred to as a heteroclinic tangle. This is a well known structural phenomenon from periodically perturbed Hamiltonian systems as described for instance in \mbox{\cite{mackay_transport_1984, ghendrih_theoretical_1996, abdullaev_asymptotical_1999, meiss_thirty_2015}}.} Previous work in tokamaks has shown the impact of the finger structure formed by the homoclinic and heteroclinic tangles which determine the heat and particle flux striation patterns discussed in \cite{evans_experimental_2005, wingen_traces_2007, schmitz_aspects_2008, frerichs_impact_2012}. 

The link between the open chaotic edge structure presented here and the $\stkout{\iota}$ values at the plasma edge has been discussed in \cite{jakubowski_modelling_2004, finken_structure_2005, textor_team_change_2006} for the cylindrical tokamak configuration at the TEXTOR-DED experiment. In 
\cite{eich_two_2000, schmitz_identification_2008, the_diii-d_and_textor_research_teams_resonant_2009} it was \hb{observed} that the finger structures that are formed by the intersecting heteroclinic tangles are comprised of shorter connection length magnetic flux tubes. Similar effects for such chaotic systems have also been \hb{found in} the chaotic boundary structure at Tore-Supra \cite{nguyen_interaction_1997, ghendrih_theoretical_1996}. 

Because of the symplectic nature \cite{abdullaev_asymptotical_1999, abdullaev_mapping_2004, finken_structure_2005} of the magnetic field in this chaotic system, these flux bundles represent local SOL channels that deposit heat and particles fluxes in the intersection zone with the wall elements. Also, in case of \hb{non-axisymmetric} plasma edge configurations, these chaotic systems are guided by magnetic turnstiles \cite{punjabi_magnetic_2022} to the wall. \hb{In this context, a turnstile represents a path for a magnetic field line to transit between regions that are otherwise not connected \mbox{\cite{mackay_transport_1984}}. These turnstiles create a pathway that brings field lines, and therefore plasma particles, from the tangles and eventually to the target. They are the reason that the escaping field lines intersect the wall in a confined helical band. The mechanism of the turnstile is discussed in detail in \mbox{\cite{punjabi_simulation_2020,punjabi_magnetic_2022}}.} We will discuss the relation of the heteroclinic tangles and the SOL flux tubes which connect to the wall via the intersection pattern in section \ref{section:flare}. \hb{The tangles and turnstiles define the behavior of the edge along with the resulting strike line pattern on the wall and therefore, the heat deposition on the PFC elements. As the edge changes, and additional chaotic structures appear, the pattern on the wall can change as well. Relating the edge chaotic structures to the changes in the wall behavior is the main focus of this paper.}

\section{Effect of Chaotic Boundary Layer on Plasma-Wall Behavior}
\label{section:flare}

\subsection{\hb{Strike-Line Analysis}}

The intersection pattern of the open chaotic boundary with the PFC is investigated in this section with strike line calculations. The PFC in this calculation is a cylindrical, toroidally symmetric wall. The minor radius of the PFC is altered to vary what aspect of the edge structure intersects the PFC. The strike points are calculated with FLARE by following a field line that is launched inside the plasma LCFS for a maximum $L_c$ of $10$ km in the forward and backward directions. Since field lines launched inside the LCFS are confined by good flux surfaces, we include a user-defined magnetic field line diffusion parameter, $d$. This is a numerical approach that allows for the field lines to enter the open chaotic edge layer, intersect the PFC, and map out a strike line pattern. For these calculations we choose the value $d=4.4\times 10^{-7} \ \text{m}^2/\text{m} $ which is approximately what would be expected from a $15$ eV electron that experiences a particle diffusion of $D=1\text{m}^2/\text{s}$. This calculation, where field lines diffuse outward from the confined region into the open chaotic system that eventually intersects the wall elements, is similar to work done on other stellarators \cite{lore_design_2014, bader_hsx_2017}. 

An overview of these field line tracing results are shown in the top plot of figure \ref{fig:spoints}. In this plot, the toroidal and poloidal angles of the intersection point of each traced field line with the PFC are recorded. Since CTH has 5-fold symmetry, only one field period is shown, spanning from toroidal angle $\phi = 0^\circ$ to $72^\circ$. Because the wall has circular cross-sections, the poloidal angle, $\theta$ is the usual cylindrical polar angle. This intersection pattern is shown in figure \ref{fig:spoints} for all toroidal currents considered, i.e. from $0$ kA to $10$ kA. 

Figure \ref{fig:spoints29} corresponds to the top plot of figure \ref{fig:spoints} and shows each toroidal current starting \hb{with the 0 kA current configuration} in the upper plot of the left column (figure \ref{fig:spoints29} A), with increasing $I_p$ going downward, starting again in the right column to end at $10$ kA in the lower right column (figure \ref{fig:spoints29} F). The upper left section of strike points and the lower right region are exactly symmetric due to stellarator symmetry, with forward/backward directions inverted. 

\hb{This strike point calculation follows the field lines for a maximum of 10 km. Because of the randomness in the diffusive property, approximately 60\% of the launched field lines strike the wall in the 0 kA configuration. Meanwhile 90\% of the points intersect the wall in the higher current cases. The difference in the number of wall-intersecting points between the low and high current case can be understood by considering that lower $I_p$ configurations have a larger confined volume. This allows for more particles to remain in the confined region.} The top plot of figure \ref{fig:spoints} indicates that the general shape of the strike point pattern is resilient in terms of being a helical pattern with a narrow spread, ranging from just a few degrees poloidally for the $0$ kA case to up to $20$ degrees for $I_p=10$ kA. This occurs despite the drastically changing magnetic structure of the open chaotic layer as shown in figure \ref{fig:poincare}. However, detailed analysis of these initial strike line calculations show that the intersection pattern can move along the helical strike line and also perpendicular to it. 

\subsection{\hb{Varying Wall Position}}

In addition to changing the current, it is also possible to alter the wall position. We study the impact of a different radius for the symmetric, cylindrical wall on the strike line pattern. In the bottom plot of figure \ref{fig:spoints}, the same calculation as the top plot\ref{fig:spoints} is shown, but with the wall moved inwards by $4$ cm. The wall is now a circular torus with minor radius of $25$ cm instead of a circular torus with a minor radius of $29$ cm. In figure \ref{fig:poincare}, the boundary at $25$ cm is plotted as the innermost green circle. The boundary at $29$ cm is just outside of the connection length contour map and is plotted in green. A similar characteristic strike line pattern with a dependency on the current levels is seen with a wall at $25$ cm compared to $29$ cm. The strike points remain in the same general area in toroidal and poloidal angles and the helical movement as well as the spread in the poloidal angle perpendicular to the helical pattern is comparable. However, there are some differences that result from the wall position. To explore the features of the strike lines in more detail, we compare the individual results for each strike point location at both radii, shown in figures \ref{fig:spoints29} and \ref{fig:spoints25}. 

Starting with the low current configurations at the innermost wall position (figure \ref{fig:spoints25}), the strike points from lower $I_p$ cases are split into two separate and distinct regions as seen in the $0$ kA and $2$ kA cases (figure \ref{fig:spoints25} A and B). In the $0$ kA case, one region exists from toroidal angle $0^\circ$ to about $20^\circ$, with another region appearing from about $50^\circ$ to $72^\circ$. As the current increases to $2$ kA, the regions move towards the center of the field period, and at $4$ kA (figure \ref{fig:spoints25} C) the two regions are connected, and they remain that way for the rest of the higher $I_p$ cases. 

\begin{figure*}
    \centering
    \includegraphics[scale=1.75]{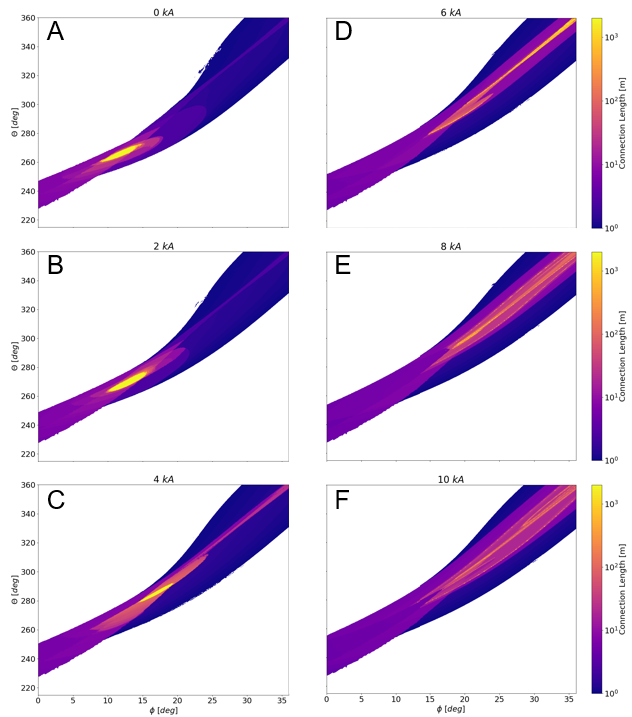}
    \caption{Connection length sampled on target at $r=25$ cm for 6 different $I_p$.}
    \label{fig:magfoot25log1}
\end{figure*}

\begin{figure*}
    \centering
    \includegraphics[scale=1.75]{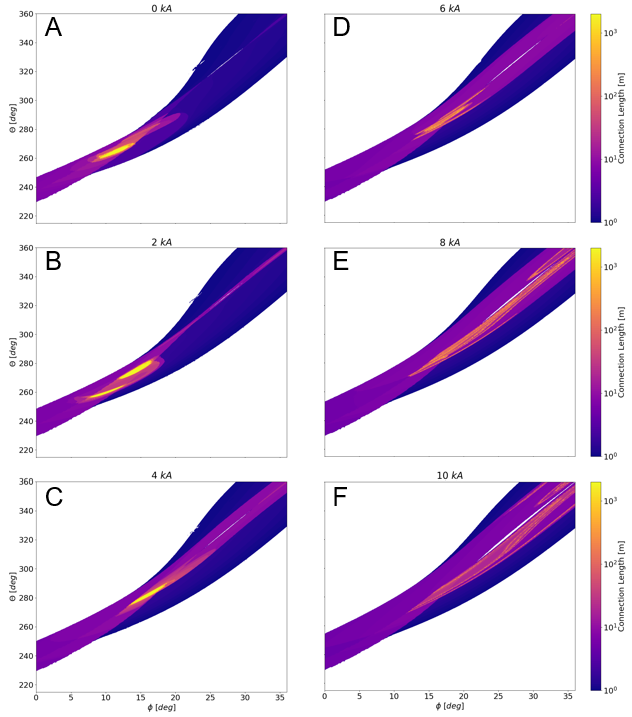}
    \caption{Connection length sampled on target at $r=27$ cm for 6 different $I_p$.}
    \label{fig:magfoot27log1}
\end{figure*}

\begin{figure*}
    \centering
    \includegraphics[scale=1.75]{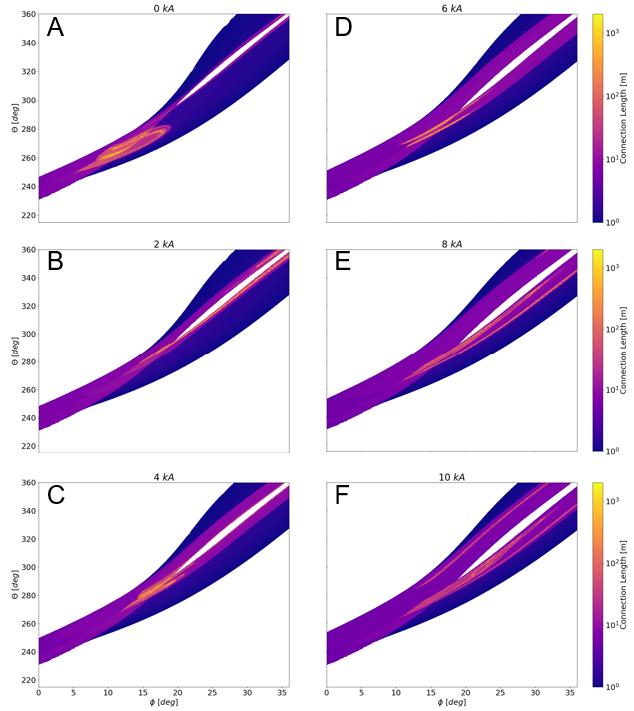}
    \caption{Connection length sampled on target at $r=29$ cm for 6 different $I_p$.}
    \label{fig:magfoot29log1}
\end{figure*}

Another metric to analyze the strike line pattern is the way field lines traced in the forward and backward toroidal direction intersect the wall in the strike line pattern. This is of interest because particles will follow these directions and their direction is indicative of the geometry and deposition pattern of later particle fluxes. Because these fluxes can yield a momentum exchange between adjacent flux tubes, understanding their geometry is important for later high density and detachment scenarios. In figure \ref{fig:spoints25}, these two directions are separated in color and marker (forward is plotted as blue $\times$ and backward as orange $\circ$). In the $0$ kA case, the separation of the areas in which the forward and backward traced field line intersect the target extends approximately $20^{\circ}$ in the poloidal direction. As the current increases, the separating line between the intersection areas from both tracing directions begins to extend more toroidally. Increasing  the plasma current increases this separation further. The forward and backward strike point regions develop a helical line of separation, and at the highest current, $10$ kA (figure \ref{fig:spoints25} F), these two regions begin to separate entirely from each other. 

This characteristic intersection pattern and its changes are comparable when the wall is moved further out as seen in figure \ref{fig:spoints29}. However, at this radius, even at the $0$ kA case (figure \ref{fig:spoints29} A), an almost complete toroidal continuation between the two forward and two backward strike regions is visible, which was not seen in the $0$ kA case in figure \ref{fig:spoints25} A for $r=25$ cm. At $2$ kA (figure \ref{fig:spoints29} B) these forward and backward regions are already completely toroidally connected. In addition, these regions are separated mostly by a helical line already in the $0$ kA (figure \ref{fig:spoints29} A) case similarly to the higher $I_p$ cases on the innermost wall. At $8$ kA (figure \ref{fig:spoints29} E), the forward and backwards strike patterns are completely helically separated. It is important to note, that the separated strike point patterns at $I_p$ = $8$ and $10$ kA are correlated with the appearance of finger-like structural flux bundles in the open chaotic layer as plotted in figure \ref{fig:poincare}. 

\subsection{\hb{Magnetic Footprint on Wall Targets}}

To understand the behavior of the intersection pattern of the field lines with the walls, we examine the connection length $L_c(\theta, \phi)$ maps on the walls. This is called the magnetic footprint on the wall and they are shown in figures \ref{fig:magfoot25log1}, \ref{fig:magfoot27log1}, and \ref{fig:magfoot29log1} for the same six different plasma current values $I_p$ which are labelled A-F in each of these figures. The magnetic footprint is shown as a contour plot of connection length $L_c(\theta, \phi)$. However, for this analysis, the field lines were not started from the inside and not given a diffusion parameter to populate as many escape channels as possible for recording strike points. Instead, the field lines were started from a spatially high resolution mesh of starting points constructed 1 mm inward of each respective wall boundary. As an example, the calculation mesh is at $r=24.9$ cm for the cylindrical wall boundary at $r=25$ cm which is plotted in figure \ref{fig:magfoot25log1}. For all points on the calculation mesh, the field lines are followed, without diffusion, for 1 km in both directions, or until they hit the wall. These plots within figure \ref{fig:magfoot25log1} each show the first half field period for $215^{\circ}< \theta <360^{\circ}$ in high resolution, which provides the details of the magnetic footprint structure. The magnetic footprints for the wall boundary of radius $r=27$ cm, an intermediate position, in figure \ref{fig:magfoot27log1}, and for $r=29$ cm in figure \ref{fig:magfoot29log1} at multiple $I_p$ are calculated and shown in the similar manner as figure \ref{fig:magfoot25log1}. The intermediate wall is shown as the intermediate green circle plotted in each Poincar\'e map in figure \ref{fig:poincare}. 

In the lower $I_p$ cases, there tends to be a localization of higher $L_c$ with lengths greater than $1$ km as seen in figure \ref{fig:magfoot25log1} as a yellow colored cluster of field lines in the $0$ kA, $2$ kA, and $4$ kA cases (figure \ref{fig:magfoot25log1} A, B, and C) with the innermost wall boundary at $r=25$ cm. In these plots, the yellow regions are predominately concentrated between $10^\circ < \phi < 20^\circ$. If $L_c$ = $2$ km, it indicates that the field line launched from the calculation mesh reaches a good flux surface surface, seen as white domain in the figure \ref{fig:poincare}. This can only happen if good flux surfaces exist within $1$ mm of the wall. Therefore, the presence of yellow regions of $2$ km connection length, which we attributed before to the confined regions, suggest a limiter rather than a divertor configuration with the open chaotic system. Instead of finger structures and short $L_c$ connection length flux bundles that represent the SOL, good flux surfaces are in direct contact with the wall. There is still a small radial domain of an open chaotic layer, and we will analyze this with 3D plasma edge transport modeling later which will show its role for impacting the heat exhaust. 

When $I_p$ is increased, we find similar trends in the magnetic footprints to the diffusive strike point calculations. The region of longest $L_c$ migrates with increasing $I_p$ to higher $\phi$ toroidally and towards the outboard mid-plane, which is defined at $\theta = 360^{\circ}$, poloidally. Furthermore, the high-resolution magnetic footprint pattern exposes a higher level of complexity, corresponding to the complexity of the open chaotic boundary layer that intersects the wall. First, at $I_p>6$ kA, no regions exist of $L_c=2$ km which supports that no good flux surface directly touch the walls. Therefore, these configurations are not limited but rather are diverted. This is congruent with the fact that based on figure \ref{fig:poincare} we identified a larger radial domain covered by field lines with finite $L_c$ and characteristic magnetic structures associated with the magnetic island chains and the invariant manifolds that form the skeleton of this divertor structure. It can be seen that the impact of increasing $I_p$ is that the areas of highest $L_c$ become less and less localized and shift to regions where the $L_c$ sampled is diverted between roughly $20^{\circ} < \phi < 36^{\circ}$. 

\hb{In the diverted region just mentioned, } the appearance of these divertor strike lines becomes increasingly evident as the boundary is placed radially outward for this calculation. This is demonstrated in  figures \ref{fig:magfoot27log1} and \ref{fig:magfoot29log1}. The splitting between the divertor strike lines increases with increasing radius at these $I_p$ values and even a private flux region can be identified which separates these divertor legs. The private flux regions appear as a narrow white region for $\phi > 20^\circ$ and $\theta > 300^\circ$. This feature is most present at higher $I_p$ and even $I_p=0$ kA case begins to display this feature at the $27$ cm boundary between $25^{\circ} < \phi < 30^{\circ}$ and $320^{\circ} < \theta < 340^{\circ}$ in figure \ref{fig:magfoot27log1}. In figure \ref{fig:magfoot29log1}, all $I_p$ cases have these divertor strike lines present.

\subsection{\hb{Transition from Limited to Diverted Edge Behavior}}

\hb{The separation of these strike lines can be seen in figure \mbox{\ref{fig:spoints29}} where the different directions of the strike points are plotted in blue and orange. By comparing the strike line figures with the magnetic footprint plots, it can be correlated that these divertor legs are indicative of the separation of the forward and backward strike points. As $I_p$ is increased, the higher yellow $L_c$ regions migrate to these divertor strike lines, as mentioned in the previous paragraph. These strike lines are the intersection point of the divertor legs with the wall as described before. The magnetic flux bundles corresponding to the finger-like structures in figure \mbox{\ref{fig:poincare}} are seen to be intercepted by the wall structures. This strongly supports a transition between a limited to diverted configuration based on the separation of these flux bundles depicted in these figures, especially for regions for $\phi > 20^{\circ}$ for the more radially outward walls. }

The calculations of strike points using field line following and the connection lengths on the walls and in the plasma volume all point to a complicated evolving edge behavior in these CTH equilibria. We find clear evidence for a limited plasma configuration and a diverted one. The most clear indication of a limited situation is a region of infinite connection length slightly in front of the wall. These are indicated by the localized yellow regions on the connection length plots in figures \ref{fig:magfoot25log1} A-C and \ref{fig:magfoot27log1} A-C \hb{which correspond to the localized strike points in figure \mbox{\ref{fig:spoints25}} A-C}. 

\hb{The} diverted plasma configurations feature the appearance of a private flux region on the wall connection length, and a splitting of the strike line pattern in the forward and backward directions as seen in \hb{figures \mbox{\ref{fig:spoints29}} A-F and } \ref{fig:magfoot29log1} A-F. \hb{In these regions from approximately $20^{\circ} < \phi < 36^{\circ}$, the long $L_c$ field lines stretch into long and thin bands of field lines surrounded by much shorter $L_c$ field lines. This behavior is analogous to what is seen in classical divertors, where we find infinite $L_c$ right at the separatrix intersection and rapidly decaying $L_c$ values outward from the separatrix into the SOL. Therefore, we refer to these elongated long $L_c$ regions that are seen for high $I_p$ values as divertor strike lines. In these configurations, the heteroclinic tangles are playing the role that is played by the divertor legs in a classical tokamak.} 

\hb{In between the limited and clearly diverted configurations, the interpretation is more difficult}. Regions of long connection lengths appear, but these do not form into coherent regions. These regions first appear consistently with a breaking of the edge flux surfaces, often resulting in what appear to be finger-like structures as seen in figure \ref{fig:poincare}. These fine structures, although visible in both the wall and volume connection length plots, tend to be smeared out by diffusive properties in both the calculations of strike points from diffusive field line following and fluid plasma simulations. 

Figures \ref{fig:magfoot25log1}, \ref{fig:magfoot27log1}, and \ref{fig:magfoot29log1} along with the strike point figures \ref{fig:spoints25} and \ref{fig:spoints29} demonstrate the variation in behavior in the magnetic topology in the edge region. We have shown the presence of a chaotic edge and demonstrated that this edge structure evolves with varying $I_p$. Furthermore, higher rotational transform $\stkout{\iota}$, associated with higher $I_p$ in figure \ref{fig:iotabar}, in turn increases the width of the chaotic layer in the edge region and the confined plasma shrinks, as seen in figure \ref{fig:poincare}. These strike points and the magnetic footprint calculations can serve as a proxy for the expected heat and particle fluxes in these regions as it has been shown in previous work that the magnetic topology is correlated to these expected fluxes \cite{abdullaev_fractal_2001, wingen_traces_2007}. 

\section{Heat Flux Calculations}
\label{section:emc3}

The analysis of the magnetic edge structure prompts questions about the transport within this open chaotic edge layer. This will be addressed using the EMC3-EIRENE 3D plasma edge fluid and kinetic neutral transport code. These are computationally expensive calculations and, therefore, were restricted to some selected configurations. The results from this edge transport modeling are compared to the field line tracing results. 

EMC3-EIRENE is a fully 3D plasma edge fluid Monte-Carlo code coupled to a 3D Monte-Carlo kinetic neutral transport model \cite{feng_3d_2004, reiter_eirene_2005}. It is the standard edge code used in stellarator research \cite{akerson_three-dimensional_2016, kawamura_three-dimensional_2018, matoike_first_2019, winters_emc3-eirene_2021} and is also commonly used in nominally axisymmetric systems that include 3D perturbations, such as tokamaks under the influence of resonant magnetic perturbations \cite{schmitz_three-dimensional_2016, dai_impacts_2020,bock_comparison_2021, frerichs_volumetric_2021}. For a full treatment of the EMC3 equations and the terms in them, please see \cite{feng_3d_2004}. 

\begin{figure*}
    \centering
    \includegraphics[scale=0.44]{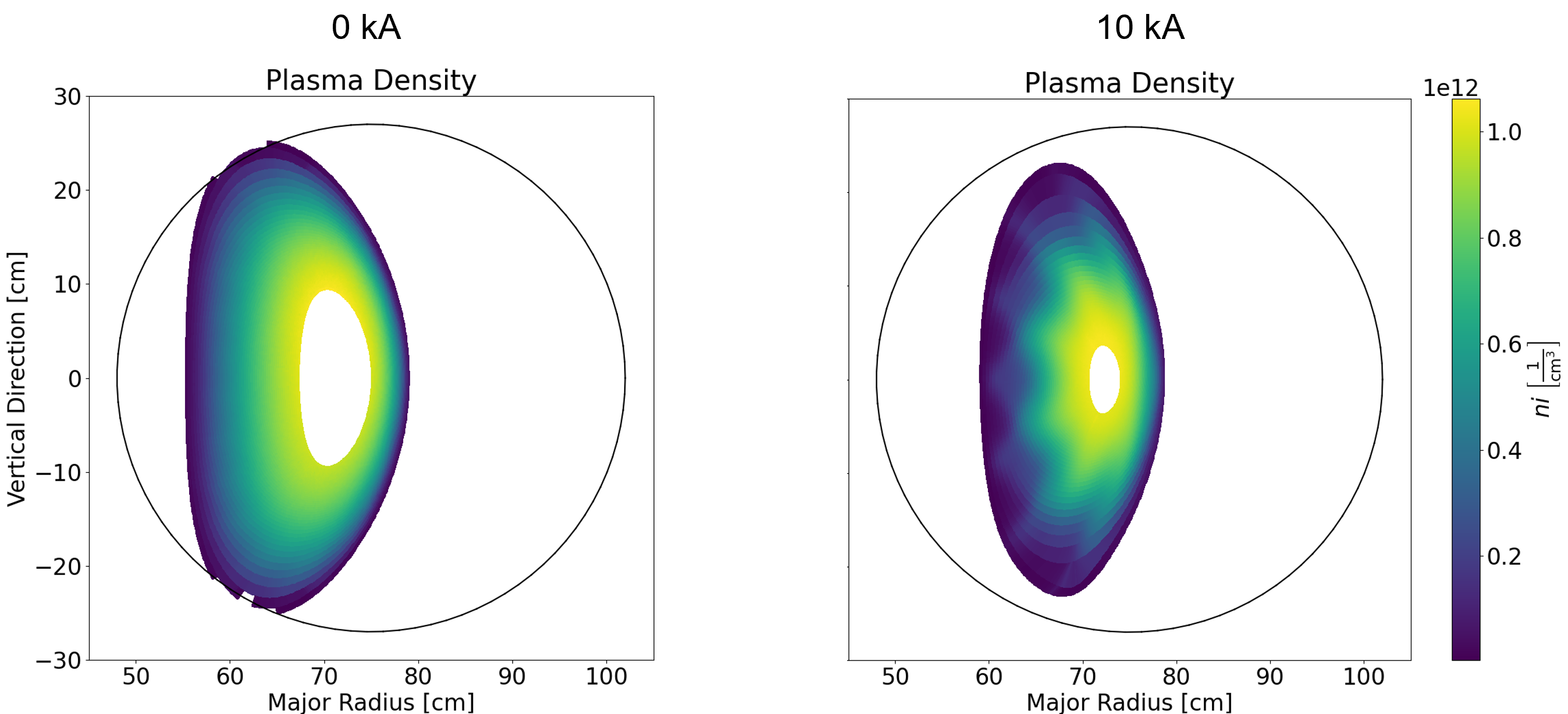}
    \caption{Plasma density $n_i$ for 0 kA (left) and 10 kA (right) with wall target at $r=$ 27 cm.}
    \label{fig:ni}
\end{figure*}

\begin{figure*}
    \centering
    \includegraphics[scale=0.44]{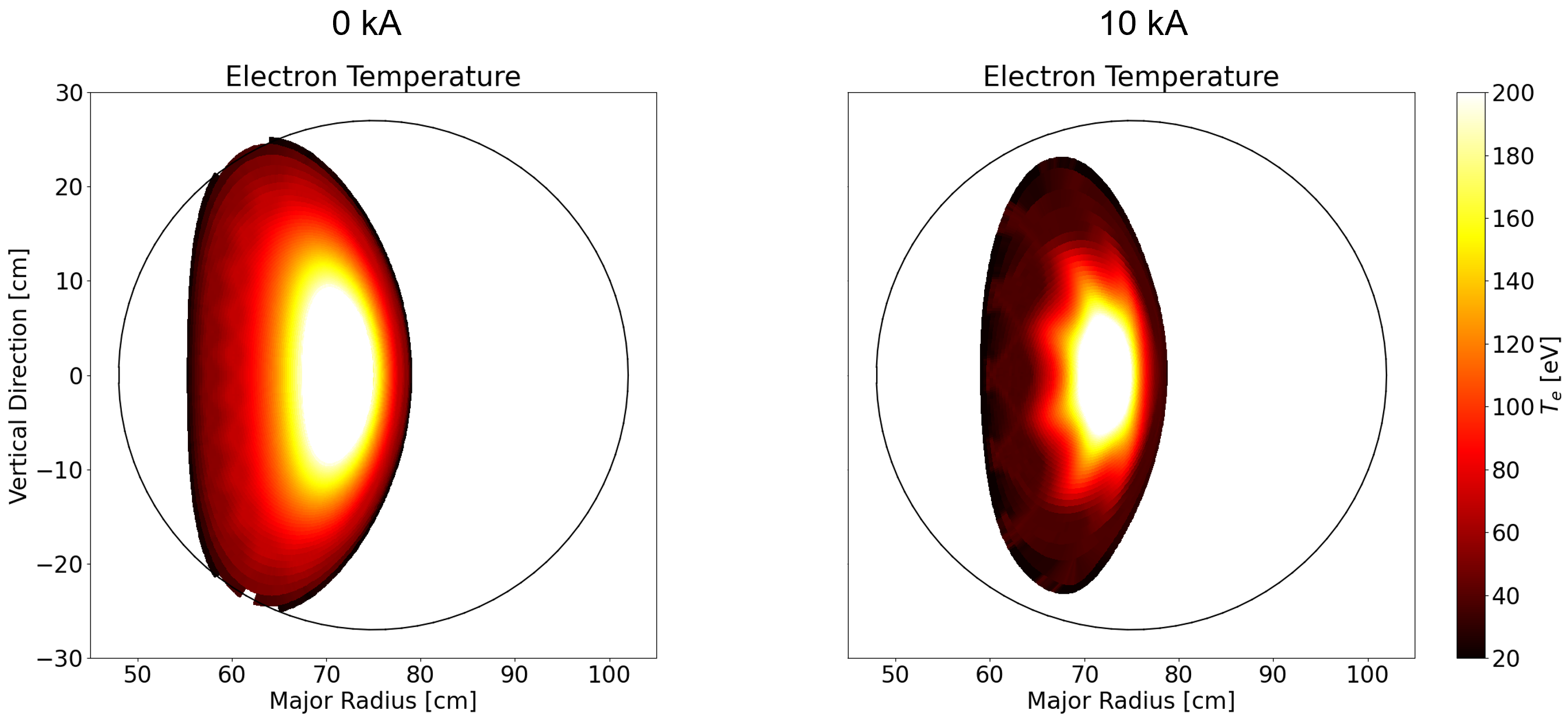}
    \caption{Electron temperature $T_e$ for 0 kA (left) and 10 kA (right) with wall target at $r=$ 27 cm.}
    \label{fig:Te}
\end{figure*}

\begin{figure*}
    \centering
    \includegraphics[scale=0.44]{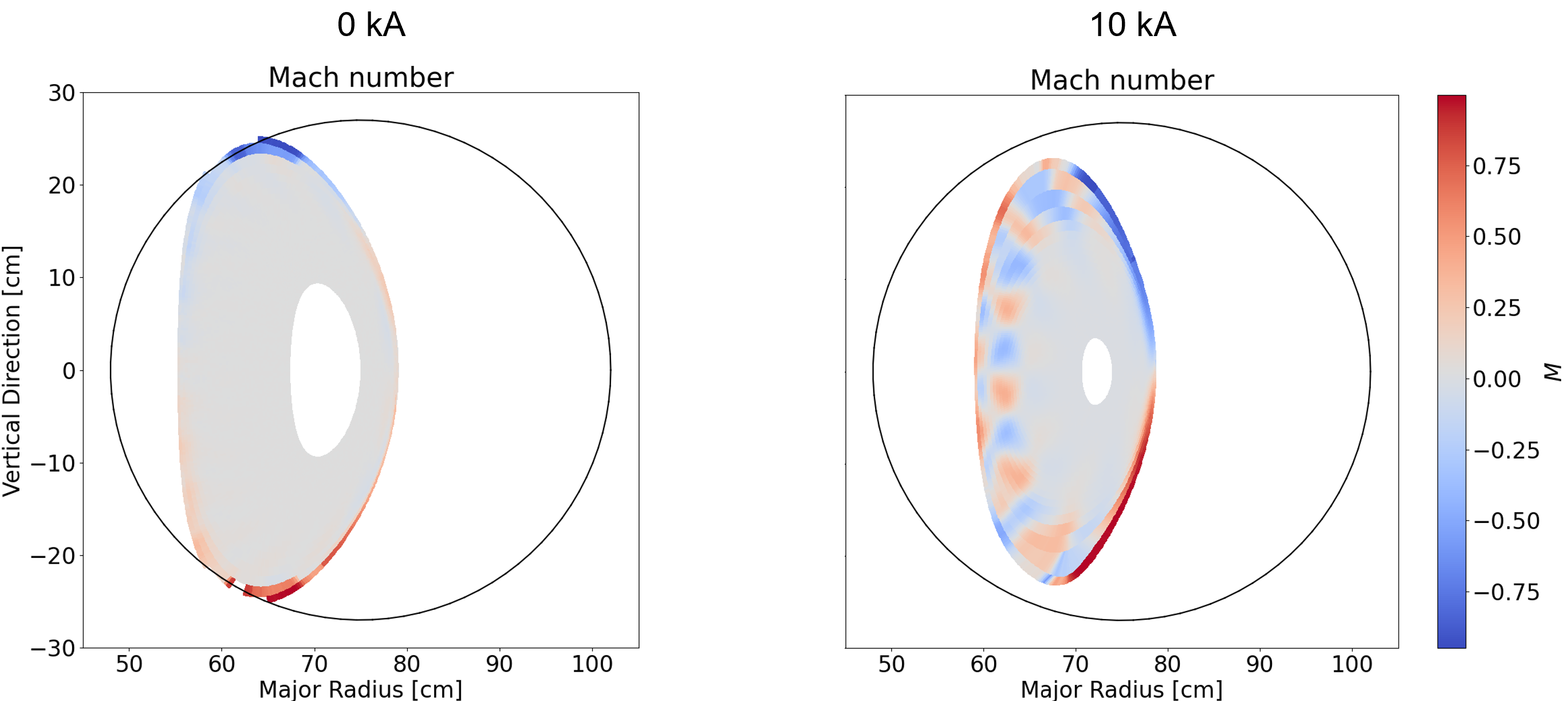}
    \caption{Mach number $M$ for 0 kA (left) and 10 kA (right) with wall target at $r=$ 27 cm.}
    \label{fig:M}
\end{figure*}

The input parameters for this analysis are as follows: The total input heating power is set to $5$ kW along with an upstream density set to $3 \times 10^{18} $ m$^{-3}$. The perpendicular transport parameters $D_i$, $\chi_i$, and $\chi_e$ represent the cross-field particle diffusion, the thermal diffusivity for ions and the thermal diffusivity for electrons, respectively. These values are not known for CTH, so we choose characteristic values of $D_i = 1 \ \text{m}^2/\text{s}$ and $\chi_i=\chi_e=3 \ \text{m}^2/\text{s}$. No impurities are included in this modeling. In figures \ref{fig:ni}, \ref{fig:Te}, \ref{fig:M}, the plasma density, electron temperature, and Mach number, respectively, are shown for the $0$ kA and $10$ kA $I_p$ cases with the wall target at $27$ cm plotted in black. These profiles reveal that the $0$ kA case has a very moderately corrugated plasma edge. This can be seen as the slight alteration in the small magnetic islands which are visible in the electron temperature in \ref{fig:Te} and the plasma flow approaches the intersection point predominantly from one side in \ref{fig:M}. This result strongly supports that the plasma in the $0$ kA case is in a limited configuration, because there is no significant radial distance between the target and the good flux surfaces which poloidally have quite homogeneous plasma parameters in close proximity to the intersection point. In contrast, the $10$ kA case seen in figure \ref{fig:M} features a significant radial domain where distinct parallel flow channels exist which separate the intersection points on the target from the remaining confined plasma region. The same is seen also in the electron temperature in figure \ref{fig:Te} where a radially extended cold region is seen and is also observed in the electron density in figure \ref{fig:ni} where this region features much reduced density values. Last but not least, the Mach number shows the evolution of independent flux tubes which feature flows of opposite sign. This clearly supports the existence of independent convective transport channels that carry the plasma flow through this radial domain. These results, therefore, strongly support that the low current configuration is limited meanwhile the higher current case of $10$ kA is diverted. 

\begin{figure*}
    \centering
    \includegraphics[scale=0.75]{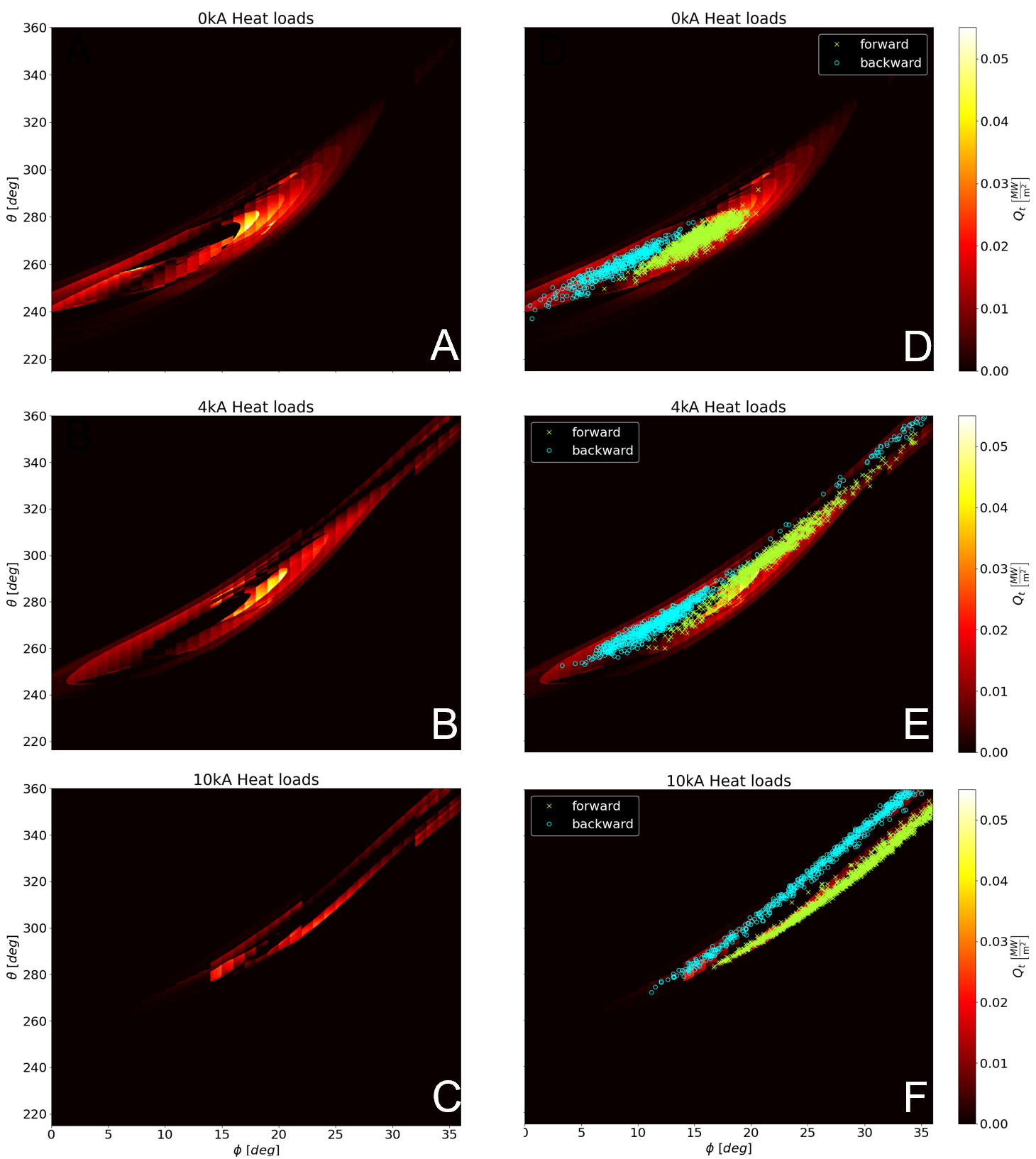}
    \caption{The left column shows EMC3-EIRENE generated heat flux deposition for a single half period on a high resolution wall target at $r=25$ cm for 0 kA (top row), 4 kA (middle row), and 10 kA (bottom row). The right column shows the same plot as the left but with FLARE strike points superimposed for the same target and with diffusion value $d=4.4\times 10^{-7} \text{m}^2/\text{m}$. }
    \label{fig:emc3_wall25}
\end{figure*}

\begin{figure*}
    \centering
    \includegraphics[scale=0.75]{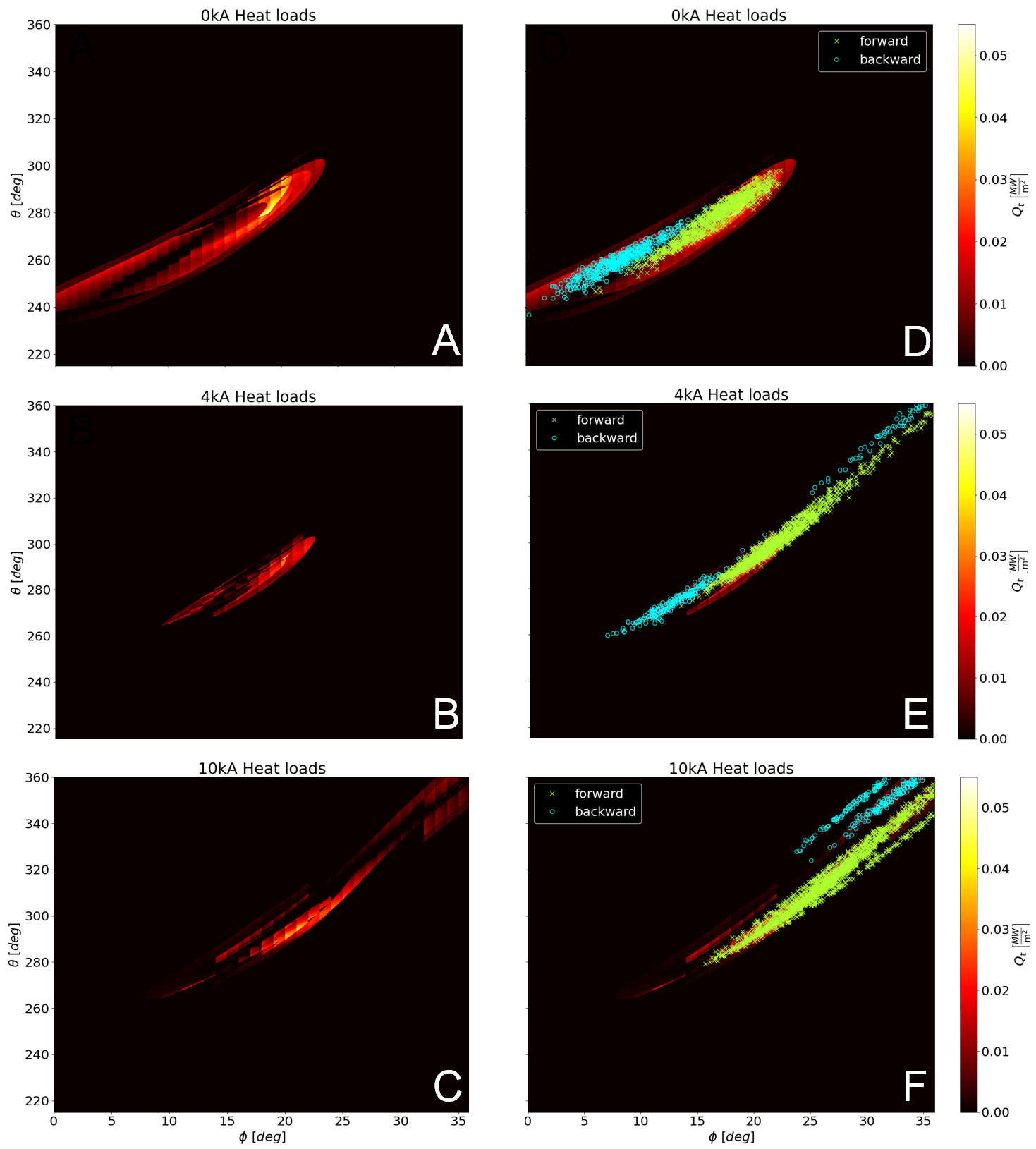}
    \caption{The left column shows EMC3-EIRENE generated heat flux deposition for a single half period on a high resolution wall target at $r=27$ cm for 0 kA (top row), 4 kA (middle row), and 10 kA (bottom row). The right column shows the same plot as the left but with FLARE strike points superimposed for the same target and with diffusion value $d=4.4\times 10^{-7} \text{m}^2/\text{m}$. }
    \label{fig:emc3_wall27}
\end{figure*}

After this analysis of the volumetric plasma characteristic in these perturbed edges, the target heat flux is analyzed using post-processing routines in EMC3-EIRENE for the $0$, $4$ and $10$ kA current configurations on the two wall positions at $r= 25$ cm and $r= 27$ cm for each $I_p$. These results are depicted in figures \ref{fig:emc3_wall25} and \ref{fig:emc3_wall27}. In each figure, the heat flux $Q_t$ on the wall calculated by EMC3-EIRENE on the left, and an overlay of the strike points for that current case from the field line diffusion calculation using FLARE on the right is shown. The forward direction of the field line is shown in green $\times$ and the backward direction in blue $\circ$, such that the points are more visible on the overlay plot. 

The EMC3-EIRENE results capture all the basic features of our previous calculations, with the exception of the $4$ kA case where the wall is at $r= 27$ cm (figure \ref{fig:emc3_wall27} B), which will be discussed later. In particular, the EMC3-EIRENE heat flux calculations indicate that the migration of the heat flux pattern in the low current, i.e. $0$ kA, configurations does not extend to $\phi = $ 36$^\circ$. Meanwhile, the heat flux deposition at high current, i.e. $10$ kA, configurations is absent between roughly 0$^\circ$ and 10 $^\circ$ in the toroidal direction and tend to extend toward $\phi=36^\circ$. In addition, the forward and backward strike points can be seen to correspond to the two major separated areas of localized heat flux for all the cases considered. In between the forward and reversed heat flux areas, the separated region of no flux eventually changes and becomes a toroidally elongated narrow band. This is most visible in the $10$ kA simulation (figure \ref{fig:emc3_wall25} C and figure \ref{fig:emc3_wall27} C). In short EMC3-EIRENE verifies the major conclusions of the previous sections including capturing the transition from a limited configuration (figures \ref{fig:emc3_wall25} A and D  and figures \ref{fig:emc3_wall27} A and D) to an intermediate configuration (figures \ref{fig:emc3_wall25} B and E and figures \ref{fig:emc3_wall27} B and E) and finally to a diverted configuration (figures \ref{fig:emc3_wall25} C and F and figures \ref{fig:emc3_wall27} C and F). 

\begin{figure*}
    \centering
    \includegraphics[scale=0.75]{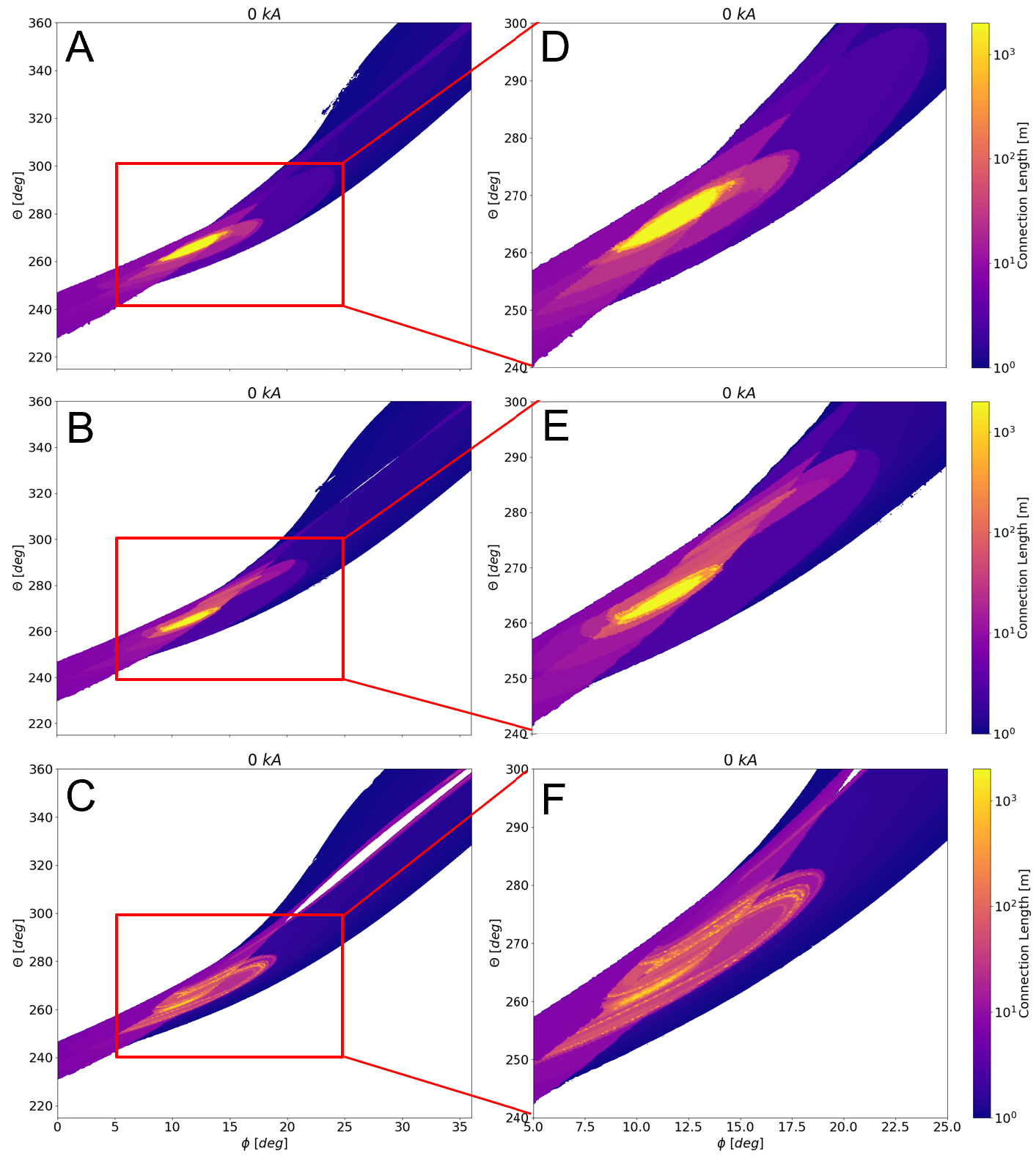}
    \caption{Magnified magnetic footprint maps of figures \ref{fig:magfoot25log1} A (shown in top row A and D), \ref{fig:magfoot27log1} A (shown in middle row B and E), and \ref{fig:magfoot29log1} A (shown in bottom row C and F).}
    \label{fig:0kA_boomerangs}
\end{figure*}

\begin{figure*}
    \centering
    \includegraphics[scale=0.5]{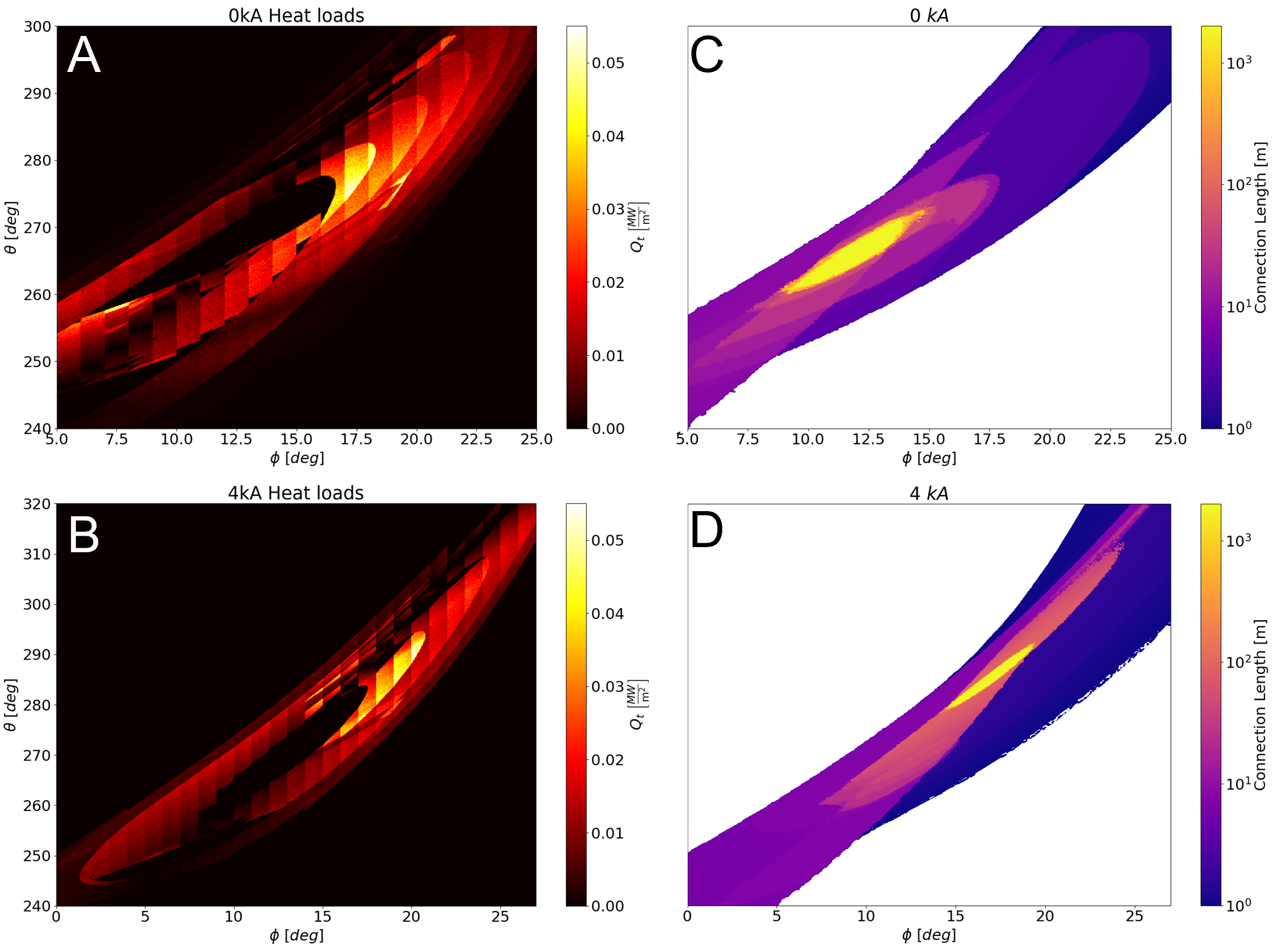}
    \caption{Plot A is a magnification of figure \ref{fig:emc3_wall25} A, plot B of figure \ref{fig:emc3_wall25} B, plot C of figure \ref{fig:magfoot25log1} A, and plot D of figure \ref{fig:magfoot25log1} B.}
    \label{fig:boomerangs25}
\end{figure*}

Looking deeper into the heat flux calculations and the magnetic footprint results, these figures show substructures which indicate a behavior correlated with homoclinic tangles mentioned in section \ref{section:cthedge}. Figure \ref{fig:0kA_boomerangs} displays the different wall boundaries (A is $r=25$ cm, B is $r=27$ cm, and C is $r=29$ cm) for the 0 kA case and a magnification of the region $240^\circ \leq \theta \leq 300^\circ$ and $5^\circ \leq \phi \leq 25^\circ$ for each magnetic footprint calculation in plots D, E, and F of figure \ref{fig:0kA_boomerangs}, respectively. By placing these alongside each other, the evolution of the substructures can be seen as the wall target configuration changes from one that is limited to one that is more diverted for a single current case. In the EMC3-EIRENE results, the two peaked heat flux regions in the current cases of 0 kA and 4 kA are comprised of substructures which appear to be boomerang-shaped lobes associated with the open chaotic layer. Plots A and B of figure \ref{fig:emc3_wall25} are magnified in figure \ref{fig:boomerangs25} A and B for the results on the wall located at $r=25$ cm for 0 kA and 4 kA, respectively. Plots C and D of figure \ref{fig:boomerangs25} similarly magnify plots A and B of figure \ref{fig:magfoot25log1} where a similarly shaped boomerang-lobe structure is seen in the magnetic footprint maps in the areas with a concentration of long $L_c$. These substructures have been previously connected to different parts of stable and unstable manifolds, and they impact the heat load deposition on the targets \cite{wingen_traces_2007}. It was observed in TEXTOR \cite{wingen_traces_2007} that there is a dependence on the field line direction which determines which manifold, stable or unstable, will receive flux. Because none of our calculations have a dependence on overall field direction, there is no expectation to see the same behavior, however, these could be observed in a dedicated experiment. Moreover, these boomerang-shaped lobes in the connection length figures tend to migrate and become helically elongated into stripes as $I_p$ increases as described in section \ref{section:flare}. The formation and migration of the helical stripes of long $L_c$ are governed by fractal patterns which are seen to impact that  heat deposition \cite{abdullaev_fractal_2001}. The details of these patterns are not resolved in the strike point behavior due to the diffusion as mentioned above. 

Turning the attention to the $4$ kA result with a wall at $r = 27$ cm (figure \ref{fig:emc3_wall27} B). In this calculation, we see the largest discrepancy between the EMC3-EIRENE heat flux calculation, which shows a very limited region of plasma wall interaction, and the strike line calculation, which shows a more elongated structure. This is in contrast to the $0$ kA result where EMC3-EIRENE shows a larger interaction region than the strike lines. There are also differences in the $10$ kA result where there are some areas where EMC3-EIRENE calculates some heat flux, but no strike lines intersect, and some areas, such as the topmost blue strike line, where EMC3-EIRENE does not predict a corresponding heat flux. These details highlight the issues associated with using simple analyses, such as the strike point calculation, instead of more advanced edge codes. 

Another note of discrepancy is the different values of diffusion used in the field line following versus the EMC3-EIRENE results. As described in section \ref{section:flare}, the value used for the diffusion is one of effectively distance corresponding to $d=4.4\times 10^{-7} \ \text{m}^2/\text{m} $ based on a 15 eV electron diffusing at 1 $\text{m}^2/\text{s}$. Meanwhile, the EMC3 simulations used characteristic diffusion of $D_i = 1 \ \text{m}^2/\text{s}$ and $\chi_i=\chi_e=3 \ \text{m}^2/\text{s}$ and, as we can see in figure \ref{fig:Te}, there is a range of values in the edge for $T_e$. Therefore, these values inherently do not coincide with one another. 

In spite of the noted differences comparing these analyses, the basic features that indicate an evolution of the strike points from a limited to diverted regime exist in both the EMC3 and strike point calculations.

\section{Discussion and Conclusion}
\label{section:disc}

\hb{The results shown here add important fundamental features to the concept of NRDs. We show that an open chaotic edge layer emerges during the toroidal current scan in CTH that is accompanied with the change from a limited to a diverted configuration. The strike line pattern stays within a general helical pattern.  Yet, differences are seen within the strike line structure. These features can be linked back to the well-established formalism of homoclinic and heteroclinic tangles used for the analysis of ergodic divertors. The tangles play the role of divertor legs where their intersections with the wall are the strike line patterns.} 

\hb{The CTH analysis supports a transition from a limited to diverted edge regime. The expected heat flux deposition calculated with EMC3-EIRENE confirms the general features seen from field line following and that these strike points can serve as a proxy for overall expected heat flux distribution on the wall targets. In particular, it is seen that the open chaotic edge for diverted cases features independent parallel flow channels that separate the target PFC from the core plasma.} 

\hb{Concerning the concept of `resiliency' in NRDs, first, there is a general resiliency of the overall intersection envelope of field lines on the wall. There exists a helical envelope that contains the strike lines for all configurations.} However, the detailed intersection of the open chaotic structure that was identified within this helical pattern changes with toroidal current. For limited cases the intersection is localized while for higher current cases, that were classified as diverted, the strike line stretches out along the target, but stays within the overall helical \hb{envelope}. This suggests that divertor target plates can be designed \hb{with respect to the} helical envelope.

The work presented here is confined to a description of the edge structure and does not attempt to answer some of the key operational questions of NRDs. In resonant divertors, the islands allow for a separation of flux tubes moving towards and away from the wall. This prevents some of the momentum loss, and was predicted to provide access to reach a high-recycling regime \cite{feng_review_2022}. The performance of the divertor in W7-X has yielded some mixed results on the effectiveness of this separation \cite{feng_understanding_2021}. NRDs, in contrast, have no natural separation, and therefore may experience momentum loss from counter-streaming flows in the plasma edge, as seen for the 10 kA case. However, it is important to note that the edge structure of stellarator configurations are in general very complicated, and that numerical codes such as EMC3-EIRENE are often difficult to run without empirical results to inform the free parameters. The performance of NRDs is an open question that can best be addressed through experiments, and then understood through simulations. 

\hb{The results presented here have already guided the development of an experimental research plan for CTH. Two Langmuir probe arrays have been constructed and installed in preparation for collection ion flux measurements at the edge of CTH plasmas.  The arrays are installed in locations where the measured ion flux is expected to undergo the greatest change when the edge flux surface structure transitions from a limited LCFS to a chaotic edge boundary. The probes are placed at two different toroidal positions and one can be moved radially which allows for comparison with computation results for the the targets placed at different radial positions. Results will be be compared to the calculations presented in this paper with regard to the strike point regions. Additionally, plasma density and temperature information from the Langmuir probes can be compared to results from EMC3-EIRENE simulations.}

While it is possible to verify the edge behavior in the low-recycling regime on a device like CTH, the conditions for plasma detachment are not achievable. However, there may be enough configuration flexibility on W7-X to allow for such research. Analysis of NRDs on W7-X would provide some needed input about the performance of such devices, which in turn can be used to plan future pilot plants and reactors.  

\ack 
The authors would like to thank Robert Mackay for some useful discussions. This work was funded by the Advanced Opportunity Fellowship (AOF) through the Graduate Engineering Research Scholars (GERS) at University of Wisconsin - Madison and by the U.S. Department of Energy under grants DE-SC0014210, DE-FG02-00ER54610, and DE-SC0014529.

\bibliographystyle{iopart-num}
% % Note the spaces between the initials

\section*{References}

\bibliography{paper}

\providecommand{\newblock}{}
\begin{thebibliography}{10}
\expandafter\ifx\csname url\endcsname\relax
  \def\url#1{{\tt #1}}\fi
\expandafter\ifx\csname urlprefix\endcsname\relax\def\urlprefix{URL }\fi
\providecommand{\eprint}[2][]{\url{#2}}

\bibitem{boozer_stellarator_2015}
Boozer A~H 2015 {\em Journal of Plasma Physics\/} {\bf 81} 515810606 ISSN
  0022-3778, 1469-7807 publisher: Cambridge University Press

\bibitem{boozer_simulation_2018}
Boozer A~H and Punjabi A 2018 Simulation of stellarator divertors: {Physics} of
  {Plasmas}: {Vol} 25, {No} 9

\bibitem{punjabi_simulation_2020}
Punjabi A and Boozer A~H 2020 {\em Physics of Plasmas\/} {\bf 27} 012503 ISSN
  1070-664X publisher: American Institute of Physics

\bibitem{grigull_first_2001}
Grigull P, McCormick K, Baldzuhn J, Burhenn R, Brakel R, Ehmler H, Feng Y,
  Gadelmeier F, Giannone L, Hartmann D, Hildebrandt D, Hirsch M, Jaenicke R,
  Kisslinger J, Knauer J, König R, Kühner G, Laqua H, Naujoks D, Niedermeyer
  H, Ramasubramanian N, Rust N, Sardei F, Wagner F, Weller A, Wenzel U and Team
  t~W~A 2001 {\em Plasma Physics and Controlled Fusion\/} {\bf 43} A175 ISSN
  0741-3335

\bibitem{renner_divertor_2002}
Renner H, Boscary J, Greuner H, Grote H, Hoffmann F~W, Kisslinger J,
  Strumberger E and Mendelevitch B 2002 {\em Plasma Physics and Controlled
  Fusion\/} {\bf 44} 1005 ISSN 0741-3335

\bibitem{wolf_performance_2019}
Wolf R~C, Alonso A, Äkäslompolo S, Baldzuhn J, Beurskens M, Beidler C~D,
  Biedermann C, Bosch H~S, Bozhenkov S, Brakel R, Braune H, Brezinsek S,
  Brunner K~J, Damm H, Dinklage A, Drewelow P, Effenberg F, Feng Y, Ford O,
  Fuchert G, Gao Y, Geiger J, Grulke O, Harder N, Hartmann D, Helander P,
  Heinemann B, Hirsch M, Höfel U, Hopf C, Ida K, Isobe M, Jakubowski M~W,
  Kazakov Y~O, Killer C, Klinger T, Knauer J, König R, Krychowiak M,
  Langenberg A, Laqua H~P, Lazerson S, McNeely P, Marsen S, Marushchenko N,
  Nocentini R, Ogawa K, Orozco G, Osakabe M, Otte M, Pablant N, Pasch E, Pavone
  A, Porkolab M, Puig~Sitjes A, Rahbarnia K, Riedl R, Rust N, Scott E,
  Schilling J, Schroeder R, Stange T, von Stechow A, Strumberger E,
  Sunn~Pedersen T, Svensson J, Thomson H, Turkin Y, Vano L, Wauters T, Wurden
  G, Yoshinuma M, Zanini M, Zhang D and {the Wendelstein 7-X Team} 2019 {\em
  Physics of Plasmas\/} {\bf 26} 082504 ISSN 1070-664X

\bibitem{nuhrenberg_quasi-helically_1988}
Nührenberg J and Zille R 1988 {\em Physics Letters A\/} {\bf 129} 113--117
  ISSN 0375-9601

\bibitem{landreman_optimization_2022}
Landreman M, Buller S and Drevlak M 2022 {\em Physics of Plasmas\/} {\bf 29}
  082501 ISSN 1070-664X

\bibitem{redl_new_2021}
Redl A, Angioni C, Belli E, Sauter O, {ASDEX Upgrade Team} and {EUROfusion MST1
  Team} 2021 {\em Physics of Plasmas\/} {\bf 28} 022502 ISSN 1070-664X

\bibitem{bader_hsx_2017}
Bader A, Boozer A~H, Hegna C~C, Lazerson S~A and Schmitt J~C 2017 {\em Physics
  of Plasmas\/} {\bf 24} 032506 ISSN 1070-664X publisher: American Institute of
  Physics

\bibitem{punjabi_magnetic_2022}
Punjabi A and Boozer A~H 2022 {\em Physics of Plasmas\/} {\bf 29} 012502 ISSN
  1070-664X

\bibitem{helander_stellarator_2012}
Helander P, Beidler C~D, Bird T~M, Drevlak M, Feng Y, Hatzky R, Jenko F,
  Kleiber R, Proll J~H~E, Turkin Y and Xanthopoulos P 2012 {\em Plasma Physics
  and Controlled Fusion\/} {\bf 54} 124009 ISSN 0741-3335 publisher: IOP
  Publishing

\bibitem{strumberger_magnetic_1992}
Strumberger E 1992 {\em Nuclear Fusion\/} {\bf 32} 737 ISSN 0029-5515

\bibitem{bader_minimum_2018}
Bader A, Hegna C~C, Cianciosa M and Hartwell G~J 2018 {\em Plasma Physics and
  Controlled Fusion\/} {\bf 60} 054003 ISSN 0741-3335 publisher: IOP Publishing

\bibitem{peterson_initial_2007}
Peterson J~T, Hartwell G~J, Knowlton S~F, Hanson J, Kelly R~F and Montgomery C
  2007 {\em Journal of Fusion Energy\/} {\bf 26} 145--148 ISSN 1572-9591

\bibitem{hartwell_design_2017}
Hartwell G~J, Knowlton S~F, Hanson J~D, Ennis D~A and Maurer D~A 2017 {\em
  Fusion Science and Technology\/} {\bf 72} 76--90 ISSN 1536-1055 publisher:
  Taylor \& Francis \_eprint: https://doi.org/10.1080/15361055.2017.1291046

\bibitem{ma_determination_2018}
Ma X, Cianciosa M~R, Ennis D~A, Hanson J~D, Hartwell G~J, Herfindal J~L, Howell
  E~C, Knowlton S~F, Maurer D~A and Traverso P~J 2018 {\em Physics of
  Plasmas\/} {\bf 25} 012516 ISSN 1070-664X
  \urlprefix\url{https://doi.org/10.1063/1.5013347}

\bibitem{hirshman_momcon_1986}
Hirshman S~P and Lee D~K 1986 {\em Computer Physics Communications\/} {\bf 39}
  161--172 ISSN 0010-4655

\bibitem{cianciosa_notitle_2016}
Cianciosa M~R, Hirshman S~P and Seal S~K 2016 {\em Bulletin of the 58th Annual
  Meeting of the APS Division of Plasma Physics, GP10.00060\/} {\bf 61}

\bibitem{frerichs_notitle_2015}
Frerichs H 2015
  \urlprefix\url{https://meetings.aps.org/Meeting/DPP15/Event/251933}

\bibitem{mackay_transport_1984}
Mackay R~S, Meiss J~D and Percival I~C 1984 {\em Physica D: Nonlinear
  Phenomena\/} {\bf 13} 55--81 ISSN 0167-2789

\bibitem{ghendrih_theoretical_1996}
Ghendrih P, Grosman A and Capes H 1996 {\em Plasma Physics and Controlled
  Fusion\/} {\bf 38} 1653 ISSN 0741-3335

\bibitem{abdullaev_asymptotical_1999}
Abdullaev S~S, Finken K~H and Spatschek K~H 1999 {\em Physics of Plasmas\/}
  {\bf 6} 153--174 ISSN 1070-664X

\bibitem{meiss_thirty_2015}
Meiss J~D 2015 {\em Chaos: An Interdisciplinary Journal of Nonlinear Science\/}
  {\bf 25} 097602 ISSN 1054-1500

\bibitem{evans_experimental_2005}
Evans T~E, Roeder R~K~W, Carter J~A, Rapoport B~I, Fenstermacher M~E and
  Lasnier C~J 2005 {\em Journal of Physics: Conference Series\/} {\bf 7} 174
  ISSN 1742-6596

\bibitem{wingen_traces_2007}
Wingen A, Jakubowski M, Spatschek K~H, Abdullaev S~S, Finken K~H, Lehnen M and
  {TEXTOR team} 2007 {\em Physics of Plasmas\/} {\bf 14} 042502 ISSN 1070-664X

\bibitem{schmitz_aspects_2008}
Schmitz O, Evans T~E, Fenstermacher M~E, Frerichs H, Jakubowski M~W, Schaffer
  M~J, Wingen A, West W~P, Brooks N~H, Burrell K~H, deGrassie J~S, Feng Y,
  Finken K~H, Gohil P, Groth M, Joseph I, Lasnier C~J, Lehnen M, Leonard A~W,
  Mordijck S, Moyer R~A, Nicolai A, Osborne T~H, Reiter D, Samm U, Spatschek
  K~H, Stoschus H, Unterberg B, Unterberg E~A, Watkins J~G, Wolf R, DIII-D t
  and Teams T 2008 {\em Plasma Physics and Controlled Fusion\/} {\bf 50} 124029
  ISSN 0741-3335

\bibitem{frerichs_impact_2012}
Frerichs H, Reiter D, Schmitz O, Cahyna P, Evans T~E, Feng Y and Nardon E 2012
  {\em Physics of Plasmas\/} {\bf 19} 052507 ISSN 1070-664X

\bibitem{jakubowski_modelling_2004}
Jakubowski M~W, Abdullaev S~S, Finken K~H and Team t~T 2004 {\em Nuclear
  Fusion\/} {\bf 44} S1 ISSN 0029-5515

\bibitem{finken_structure_2005}
Finken K, Abdullaev S, Jakubowski M, Lehnen M, Nicolai A and Spatschek K 2005
  The structure of magnetic field in the {TEXTOR}-{DED} Tech. Rep. 1433-5522
  Germany iNIS-DE--0120 INIS Reference Number: 37040633

\bibitem{textor_team_change_2006}
{TEXTOR Team}, Jakubowski M~W, Schmitz O, Abdullaev S~S, Brezinsek S, Finken
  K~H, Krämer-Flecken A, Lehnen M, Samm U, Spatschek K~H, Unterberg B and Wolf
  R~C 2006 {\em Physical Review Letters\/} {\bf 96} 035004 publisher: American
  Physical Society
  \urlprefix\url{https://link.aps.org/doi/10.1103/PhysRevLett.96.035004}

\bibitem{eich_two_2000}
Eich T, Reiser D and Finken K~H 2000 {\em Nuclear Fusion\/} {\bf 40} 1757 ISSN
  0029-5515

\bibitem{schmitz_identification_2008}
Schmitz O, Jakubowski M~W, Frerichs H, Harting D, Lehnen M, Unterberg B,
  Abduallaev S~S, Brezinsek S, Classen I, Evans T, Feng Y, Finken K~H, Kantor
  M, Reiter D, Samm U, Schweer B, Sergienko G, Spakman G~W, Tokar M, Uzgel E,
  Wolf R~C and Team t~T 2008 {\em Nuclear Fusion\/} {\bf 48} 024009 ISSN
  0029-5515

\bibitem{the_diii-d_and_textor_research_teams_resonant_2009}
{the DIII-D and TEXTOR Research Teams}, Schmitz O, Evans T~E, Fenstermacher
  M~E, Unterberg E~A, Austin M~E, Bray B~D, Brooks N~H, Frerichs H, Groth M,
  Jakubowski M~W, Lasnier C~J, Lehnen M, Leonard A~W, Mordijck S, Moyer R~A,
  Osborne T~H, Reiter D, Samm U, Schaffer M~J, Unterberg B and West W~P 2009
  {\em Physical Review Letters\/} {\bf 103} 165005 publisher: American Physical
  Society

\bibitem{nguyen_interaction_1997}
Nguyen F, Chendrih P and Grosman A 1997 {\em Nuclear Fusion\/} {\bf 37} 743
  ISSN 0029-5515

\bibitem{abdullaev_mapping_2004}
Abdullaev S~S 2004 {\em Nuclear Fusion\/} {\bf 44} S12 ISSN 0029-5515

\bibitem{lore_design_2014}
Lore J~D, Andreeva T, Boscary J, Bozhenkov S, Geiger J, Harris J~H, Hoelbe H,
  Lumsdaine A, McGinnis D, Peacock A and Tipton J 2014 {\em IEEE Transactions
  on Plasma Science\/} {\bf 42} 539--544 ISSN 1939-9375 conference Name: IEEE
  Transactions on Plasma Science

\bibitem{abdullaev_fractal_2001}
Abdullaev S~S, Eich T and Finken K~H 2001 {\em Physics of Plasmas\/} {\bf 8}
  2739--2749 ISSN 1070-664X

\bibitem{feng_3d_2004}
Feng Y, Sardei F, Kisslinger J, Grigull P, McCormick K and Reiter D 2004 {\em
  Contributions to Plasma Physics\/} {\bf 44} 57--69 ISSN 1521-3986 \_eprint:
  https://onlinelibrary.wiley.com/doi/pdf/10.1002/ctpp.200410009

\bibitem{reiter_eirene_2005}
Reiter D, Baelmans M and Börner P 2005 {\em Fusion Science and Technology\/}
  {\bf 47}

\bibitem{akerson_three-dimensional_2016}
Akerson A~R, Bader A, Hegna C~C, Schmitz O, Stephey L~A, Anderson D~T, Anderson
  F~S~B and Likin K~M 2016 {\em Plasma Physics and Controlled Fusion\/} {\bf
  58} 084002 ISSN 0741-3335 publisher: IOP Publishing

\bibitem{kawamura_three-dimensional_2018}
Kawamura G, Tanaka H, Mukai K, Peterson B, Dai S~Y, Masuzaki S, Kobayashi M,
  Suzuki Y, Feng Y and Group L~E 2018 {\em Plasma Physics and Controlled
  Fusion\/} {\bf 60} 084005 ISSN 0741-3335 publisher: IOP Publishing

\bibitem{matoike_first_2019}
Matoike R, Kawamura G, Ohshima S, Kobayashi M, Suzuki Y, Nagasaki K, Masuzaki
  S, Kobayashi S, Yamamoto S, Kado S, Minami T, Okada H, Konoshima S, Mizuuchi
  T, Tanaka H, Matsuura H, Feng Y and Frerichs H 2019 {\em Plasma and Fusion
  Research\/} {\bf 14} 3403127--3403127

\bibitem{winters_emc3-eirene_2021}
Winters V~R, Reimold F, König R, Krychowiak M, Romba T, Biedermann C,
  Bozhenkov S, Drewelow P, Endler M, Feng Y, Frerichs H, Fuchert G, Geiger J,
  Gao Y, Harris J~H, Jakubowski M, Kornejew P, Kremeyer T, Niemann H, Pasch E,
  Puig-Sitjes A, Schlisio G, Scott E~R, Wurden G~A and Team t~W~X 2021 {\em
  Plasma Physics and Controlled Fusion\/} {\bf 63} 045016 ISSN 0741-3335
  publisher: IOP Publishing

\bibitem{schmitz_three-dimensional_2016}
Schmitz O, Becoulet M, Cahyna P, Evans T~E, Feng Y, Frerichs H, Loarte A, Pitts
  R~A, Reiser D, Fenstermacher M~E, Harting D, Kirschner A, Kukushkin A, Lunt
  T, Saibene G, Reiter D, Samm U and Wiesen S 2016 {\em Nuclear Fusion\/} {\bf
  56} 066008 ISSN 0029-5515 publisher: IOP Publishing

\bibitem{dai_impacts_2020}
Dai S~Y, Zhang H~M, Lyu B, Wang L, Feng Y, Wang Z~X and Wang D~Z 2020 {\em
  Journal of Plasma Physics\/} {\bf 86} 815860303 ISSN 0022-3778, 1469-7807
  publisher: Cambridge University Press

\bibitem{bock_comparison_2021}
Bock L, Brida D, Faitsch M, Schmid K, Lunt T and Team t~A~U 2021 {\em Nuclear
  Fusion\/} {\bf 62} 026020 ISSN 0029-5515 publisher: IOP Publishing

\bibitem{frerichs_volumetric_2021}
Frerichs H, Feng Y, Bonnin X, Pitts R~A, Reiter D and Schmitz O 2021 {\em
  Physics of Plasmas\/} {\bf 28} 102503 ISSN 1070-664X

\bibitem{feng_review_2022}
Feng Y and {W7-X-team} 2022 {\em Plasma Physics and Controlled Fusion\/} {\bf
  64} 125012 ISSN 0741-3335 publisher: IOP Publishing

\bibitem{feng_understanding_2021}
Feng Y, Jakubowski M, König R, Krychowiak M, Otte M, Reimold F, Reiter D,
  Schmitz O, Zhang D, Beidler C~D, Biedermann C, Bozhenkov S, Brunner K~J,
  Dinklage A, Drewelow P, Effenberg F, Endler M, Fuchert G, Gao Y, Geiger J,
  Hammond K~C, Helander P, Killer C, Knauer J, Kremeyer T, Pasch E,
  Rudischhauser L, Schlisio G, Pedersen T~S, Wenzel U, Winters V and team W~X
  2021 {\em Nuclear Fusion\/} {\bf 61} 086012 ISSN 0029-5515 publisher: IOP
  Publishing

\end{thebibliography}

\end{document}